\documentclass[longauth]{aa} 
\usepackage[varg]{txfonts}
\usepackage{graphicx}
\usepackage{subfigure}
\usepackage{float}
\usepackage{natbib,twoopt}
\usepackage{xspace}
\usepackage{balance}
\usepackage{upgreek}
\usepackage{multirow}
\usepackage{color}
\usepackage[breaklinks=true]{hyperref}

\bibpunct{(}{)}{;}{a}{}{,} %
\DeclareGraphicsExtensions{.eps,.ps,.pdf}

\begin{document}

   \title{The VIMOS Public Extragalactic Redshift Survey (VIPERS)}
   \subtitle{Hierarchical scaling and biasing
\thanks{based on
observations collected at the European Southern Observatory, Cerro Paranal, 
Chile, using the Very Large Telescope under 
programs 182.A-0886 and partly 070.A-9007.
Also based on observations obtained with MegaPrime/MegaCam, a joint project of 
CFHT and CEA/DAPNIA, at the 
Canada-France-Hawaii Telescope (CFHT), which is operated by the
National Research Council (NRC) of Canada, the Institut National des Sciences 
de l’Univers of the Centre National de la 
Recherche Scientifique (CNRS) of France, and the University of Hawaii. This 
work is based in part on data products produced 
at TERAPIX and the Canadian Astronomy Data Centre as part of the 
Canada-France-Hawaii Telescope Legacy Survey, 
a collaborative project of NRC and CNRS. The VIPERS web site is 
http://www.vipers.inaf.it.}
}
\author{
A. Cappi\inst{1,2}
\and F.~Marulli\inst{3,4,1}
\and J.~Bel\inst{5}
\and O.~Cucciati\inst{3,1}    
\and E.~Branchini\inst{6,7,8}
\and S.~de la Torre\inst{9}
\and L.~Moscardini\inst{3,4,1}
\and M.~Bolzonella\inst{1}
\and L.~Guzzo\inst{10,11}
\and U.~Abbas\inst{12}
\and C.~Adami\inst{9}
\and S.~Arnouts\inst{13}
\and D.~Bottini\inst{14}
\and J.~Coupon\inst{15}           
\and I.~Davidzon\inst{3,1}
\and G.~De Lucia\inst{16}
\and A.~Fritz\inst{14}
\and P.~Franzetti\inst{14}
\and M.~Fumana\inst{14}
\and B.~Garilli\inst{14,9}     
\and B.~R.~Granett\inst{6}
\and O.~Ilbert\inst{9}
\and A.~Iovino\inst{10}
\and J.~Krywult\inst{17}
\and V.~Le Brun\inst{9}
\and O.~Le F\`evre\inst{9}
\and D.~Maccagni\inst{14}
\and K.~Ma{\l}ek\inst{18,19}
\and H.~J.~McCracken\inst{20}
\and L.~Paioro\inst{14}
\and M.~Polletta\inst{14}
\and A.~Pollo\inst{21,19}
\and M.~Scodeggio\inst{14}
\and L.~A.~.M.~Tasca\inst{9}
\and R.~Tojeiro\inst{22}
\and D.~Vergani\inst{23,1}
\and A.~Zanichelli\inst{24}
\and A.~Burden\inst{22}       
\and C.~Di Porto\inst{1}      
\and A.~Marchetti\inst{25,10} 
\and C.~Marinoni\inst{5}      
\and Y.~Mellier\inst{20}      
\and R.~C.~Nichol\inst{22}    
\and J.~A.~Peacock\inst{26}   
\and W.~J.~Percival\inst{22}  
\and S.~Phleps\inst{27}
\and C.~Schimd\inst{9}
\and H.~Schlagenhaufer\inst{28,27}
\and M.~Wolk\inst{20}         
\and G.~Zamorani\inst{1}      
}
\offprints{A. Cappi \\ \email{alberto.cappi@oabo.inaf.it}}

\institute{
INAF - Osservatorio Astronomico di Bologna, via Ranzani 1, I-40127, Bologna, Italy 
\and Laboratoire Lagrange, UMR7293, Universit\'e de Nice Sophia Antipolis, CNRS, Observatoire de la C\^ote d’Azur, 06300 Nice, France 
\and Dipartimento di Fisica e Astronomia - Alma Mater Studiorum Universit\`{a} di Bologna, viale Berti Pichat 6/2, I-40127 Bologna, Italy 
\and INFN, Sezione di Bologna, viale Berti Pichat 6/2, I-40127 Bologna, Italy 
\and Centre de Physique Th\'eorique, UMR 6207 CNRS-Universit\'e de Provence, Case 907, F-13288 Marseille, France 
\and Dipartimento di Matematica e Fisica, Universit\`{a} degli Studi Roma Tre, via della Vasca Navale 84, 00146 Roma, Italy 
\and INFN, Sezione di Roma Tre, via della Vasca Navale 84, I-00146 Roma, Italy 
\and INAF - Osservatorio Astronomico di Roma, via Frascati 33, I-00040 Monte Porzio Catone (RM), Italy 
\newpage
\and Aix Marseille Universit\'e, CNRS, LAM (Laboratoire d'Astrophysique de Marseille) UMR 7326, 13388, Marseille, France  
\and INAF - Osservatorio Astronomico di Brera, Via Brera 28, 20122 Milano, via E. Bianchi 46, 23807 Merate, Italy 
\and Dipartimento di Fisica, Universit\`a di Milano-Bicocca, P.zza della Scienza 3, I-20126 Milano, Italy 
\and INAF - Osservatorio Astronomico di Torino, 10025 Pino Torinese, Italy 
\and Canada-France-Hawaii Telescope, 65--1238 Mamalahoa Highway, Kamuela, HI 96743, USA 
\and INAF - Istituto di Astrofisica Spaziale e Fisica Cosmica Milano, via Bassini 15, 20133 Milano, Italy
\and Astronomical Observatory of the  University of Geneva, ch. d'Ecogia
16, 1290 Versoix, Switzerland 
\and INAF - Osservatorio Astronomico di Trieste, via G. B. Tiepolo 11, 34143 Trieste, Italy 
\and Institute of Physics, Jan Kochanowski University, ul. Swietokrzyska 15, 25-406 Kielce, Poland 
\and Department of Particle and Astrophysical Science, Nagoya University, Furo-cho, Chikusa-ku, 464-8602 Nagoya, Japan 
\and National Centre for Nuclear Research, ul. Hoza 69, 00-681 Warszawa, Poland 
\and Institut d'Astrophysique de Paris, UMR7095 CNRS, Universit\'{e} Pierre et Marie Curie, 98 bis Boulevard Arago, 75014 Paris, 
France 
\and Astronomical Observatory of the Jagiellonian University, Orla 171, 30-001 Cracow, Poland 
\and Institute of Cosmology and Gravitation, Dennis Sciama Building, University of Portsmouth, Burnaby Road, Portsmouth, PO1 3FX 
\and INAF - Istituto di Astrofisica Spaziale e Fisica Cosmica Bologna, via Gobetti 101, I-40129 Bologna, Italy 
\and INAF - Istituto di Radioastronomia, via Gobetti 101, I-40129, Bologna, Italy 
\and Universit\`{a} degli Studi di Milano, via G. Celoria 16, 20130 Milano, Italy 
\and SUPA, Institute for Astronomy, University of Edinburgh, Royal Observatory, Blackford Hill, Edinburgh EH9 3HJ, UK 
\and Max-Planck-Institut f\"{u}r Extraterrestrische Physik, D-84571 Garching b. M\"{u}nchen, Germany 
\and Universit\"{a}tssternwarte M\"{u}nchen, Ludwig-Maximillians Universit\"{a}t, Scheinerstr. 1, D-81679 M\"{u}nchen, Germany 
}

\date{Received ..., 2015; accepted ..., 2015}

%
%

\abstract
{}
{Building on the two-point correlation function analyses of the 
VIMOS Public Extragalactic Redshift Survey (VIPERS), we investigate the 
higher-order correlation properties of the same galaxy samples to test the 
hierarchical scaling hypothesis at $z \sim 1$ and the dependence on galaxy 
luminosity, stellar mass, and redshift. 
With this work we also aim to assess possible deviations from the 
linearity of galaxy bias independently from a previously performed 
analysis of our survey. 
}
{We have measured the count probability distribution function in 
spherical cells of varying radii ($3 \le R \le 10 h^{-1}$ Mpc), 
deriving  $\sigma_{8g}$ (the galaxy rms at 8 $h^{-1}$ Mpc), 
the volume--averaged two-, three-, and four--point correlation functions
and the normalized skewness $S_{3g}$ and kurtosis $S_{4g}$
 for different volume--limited subsamples, covering the following ranges: 
$-19.5 \le M_B(z=1.1) - 5 \log(h) \le -21.0$ in absolute magnitude, 
$9.0 \le \log(M_*/M_\odot h^{-2}) \le 11.0$ in stellar mass, and
$0.5 \le z < 1.1$  in redshift.}
{We have performed the first measurement of high--order correlation functions 
at $z \sim 1$ in a spectroscopic redshift survey. Our main results are the 
following.
1) 
The hierarchical scaling between the volume--averaged two- and 
three-point and two- and four--point correlation 
functions holds throughout the whole range of scale and redshift we could test.
2) We do not find a significant dependence of 
$S_{3g}$ on luminosity (below $z=0.9$ the value of $S_{3g}$ decreases with 
luminosity, but only at $1 \sigma$--level). 
3) We do not detect a significant dependence of $S_{3g}$ and $S_{4g}$ on scale, 
except beyond $z \sim 0.9$, where $S_{3g}$ and $S_{4g}$ have higher values on 
large 
scales ($R \ge 10 h^{-1}$ Mpc): this increase is mainly due to one of the two 
CFHTLS Wide Fields observed by VIPERS and can be explained as a consequence of
sample variance, consistently with our analysis of mock catalogs.
4) We do not detect a significant evolution of $S_{3g}$ and $S_{4g}$ with 
redshift (apart from the increase of their values with scale
in the last redshift bin).
5) $\sigma_{8g}$ increases with luminosity, but does not show significant 
evolution with redshift. As a consequence, the linear bias factor 
$b=\sigma_{8g}/\sigma_{8m}$, where 
$\sigma_{8m}$ is the rms of matter at a scale of 8~$h^{-1}$ Mpc, 
increases with redshift, in agreement with 
 the independent analysis of VIPERS 
and of other surveys such as the VIMOS--VLT Deep Survey (VVDS). 
We measure the lowest bias $b=1.47 \pm 0.18$ for galaxies
 with  $M_B(z=1.1)-5\log(h) \le -19.5$  in the first redshift bin
($0.5 \le z < 0.7$) and the 
highest bias  $b=2.12 \pm 0.28$ for galaxies with 
$M_B(z=1.1)-5\log(h) \le -21.0$ in the last redshift bin 
($0.9 \le z < 1.1$). 
6) We quantify deviations from the linear bias by means of the Taylor expansion
parameter  $b_2$.
We obtain $b_2 = -0.20 \pm 0.49$ for $0.5 \le z < 0.7$ 
and  $b_2 = -0.24 \pm 0.35$ for $0.7 \le z < 0.9$, while for the 
redshift range $0.9 \le z < 1.1$ we find $b_2 = +0.78 \pm 0.82$.
These results are compatible with a null non-linear bias term,
but taking into account another analysis for VIPERS and the analysis of other 
surveys, 
we argue that there is evidence for a small but non-zero 
non-linear bias term.
}
{}


\authorrunning{Cappi et al.}
\titlerunning{VIPERS: Hierarchical Scaling and Biasing}

\maketitle

%
%

\section{Introduction}

In the standard model of structure formation,
the growth of density fluctuations from a primordial Gaussian density field 
is driven by gravity;
it is possible 
to follow the evolution of these fluctuations through analytical
and numerical approaches and predict the statistical 
properties for the dark matter field and dark matter haloes.
Galaxies form in a complex process following the 
baryonic infall into dark matter halos: this means that the comparison between
theory and observations is not straightforward, but it also implies
that the spatial distribution of galaxies contains a wealth of information 
relevant for both cosmology and the physics of galaxy formation.

Extracting and exploiting this information from the data requires a 
number of different and complementary statistical approaches. For example,
while the two--point correlation function $\xi_2({\bf r})$ is the
simplest and most widely used statistical indicator of galaxy clustering,
a complete description of a distribution is only given by
the full $J$--point correlation functions $\xi_J$, or equivalently, by 
the volume--averaged 
correlation functions $\overline{\xi}_J$, which are related to the 
$J$--order moments of 
the count probability distribution function (PDF)\footnote{However, there is 
the important exception of the lognormal distribution, see 
\citealp{1991MNRAS.248....1C} and                     
\citealp{2011ApJ...738...86C}.}.                       
The count PDF gives the probability of counting $N$ objects as a function of 
volume $V$. 
High--order correlations are particularly interesting because
perturbation theory and numerical simulations can describe their behaviour 
for the gravitational evolution of matter density fluctuations.

The first estimates of the two-- and three--point galaxy correlations 
functions on angular catalogues of galaxies were made by 
\cite{1977ApJ...217..385G},       
who found that these estimates were well described by the hierarchical 
relation 
$\xi_3(r_{12},r_{13},r_{23}) = Q [\xi_2(r_{12}) \xi_2(r_{13}) + 
\xi_2(r_{13}) \xi_2(r_{23}) +  \xi_2(r_{12}) \xi_2(r_{23})]$. 
The three--point correlation function has subsequently become a standard 
statistical tool
for the analysis of clustering and has been applied to simulations and recent
surveys of galaxies (see e.g. \citealp{2008ApJ...672..849M}, 
\citealp{2014MNRAS.443.2874M}), 
while its Fourier transform, the bispectrum,
has also been applied to the analysis of the 
Ly$_\alpha$ forest (\citealp{2003MNRAS.344..776M}, \citealp{2004MNRAS.347L..26V})
and of the cosmic microwave background (CMB) (\citealp{2013arXiv1303.5084P}). 

The scaling relation between the two-- and three--point correlation functions
was soon generalized to higher orders
(\citealp{1978ApJ...221...19F} up to $J=4$,               
\citealp{1984A&A...130...79S} up to $J=5$)               
and was mathematically described by the so--called hierarchical models, where
the $J$--point correlation functions are expressed as a function 
of products of the two--point correlation function. 
Different versions of these models were suggested, but 
\cite{1989A&A...220....1B} 
showed that all of them belong to the general class
of scale--invariant models, which are defined by the scaling property:

\begin{equation}
 \xi_J(\lambda r_1, ..., \lambda r_J) = \lambda ^{-(J-1)\gamma} \xi_J(r_1, ...,
r_J).
\end{equation}

From a physical point of view, the hierarchical scaling of the correlation
functions is expected in the highly non-linear regime 
(the BBGKY hierarchy, see 
\citealp{1977ApJS...34..425D},       
\citealp{1984ApJ...277L...5F},       
\citealp{1988ApJ...332...67H})       
and in the quasi--linear regime (from perturbation theory, 
see \citealp{1980lssu.book.....P},   
\citealp{1984ApJ...279..499F},       
\citealp{1992ApJ...392....1B},       
\citealp{2002PhR...367....1B}        
and references therein). 

Another prediction of the hierarchical models is that 
the normalized high--order reduced moments 
$S_J \equiv \overline{\xi}_J / \overline{\xi}_2 ^{J-1}$ should be constant.
In the present paper we focus on the normalized skewness 
$S_3$ and kurtosis $S_4$.
\cite{1980lssu.book.....P} 
showed that in second--order 
perturbation theory, assuming Gaussian primordial density fluctuations and an 
Einstein-de Sitter model, $S_{3m}$, the normalized skewness of matter 
fluctuations assumes the value $34/7$. 
Subsequent works have shown that the smoothed $S_{3m}$ depends 
on the slope of the power spectrum and has a very weak dependence on the 
cosmological model 
(see \citealp{2002PhR...367....1B}). 

While in standard models with Gaussian primordial fluctuations
the skewness and higher--order moments assume non--zero values 
as a consequence of gravitational clustering, scenarios with 
non--Gaussian primordial perturbations also predict a primordial non--zero 
skewness, particularly at large scales ($\ge 10 h^{-1}$ Mpc)
(\citealp{1993ApJ...408...33L}, 
\citealp{1994ApJ...429...36F}, 
\citealp{1996ApJ...462L...1G}, 
\citealp{1998MNRAS.301..524G}, 
\citealp{2000PhRvD..62b1301D}); 
therefore these scenarios can in principle be constrained by measuring 
the high--order moments (\citealp{2014MNRAS.443.1402M}). 

Moreover, it has been shown that the hierarchy of the $J$--point 
functions 
and the measurement of $S_3$ and $S_4$ can be used as a cosmological test
to distinguish between the standard $\Lambda CDM$ and models including 
long-range scalar interaction between dark matter particles (``fifth force''  
DM models), as shown by 
\cite{2010PhRvD..82j3536H}, 
who found the largest deviations in the redshift range $0.5 < z < 2$.

However, the comparison between the theoretical predictions 
for the matter distribution and the observed galaxy distribution 
is not trivial, as a consequence of bias.
One of the first results derived from the analysis of the first redshift 
surveys was that the amplitude of the two--point correlation function
depends on galaxy luminosity and galaxy colour
(see \citealp{2013A&A...557A..17M} 
and references therein); therefore, the galaxy distribution must generally 
differ from the underlying matter distribution.
A common assumption is that the galaxy and matter density fields are related by
a linear relation,  $\delta_g = b \delta_m$, where 
$\delta_g \equiv \Delta\rho_g/\rho_g$ and $\delta_m \equiv \Delta\rho_m/\rho_m$
are the galaxy and matter density contrast, respectively.
This relation is a consequence of
the scenario of biased galaxy formation, where galaxies form above
a given threshold of the linear density field, in the limit of high threshold
and low variance. Of course, this relation cannot have general validity: 
when $b>1$ and $\delta_m < 0$, the linear relation gives an unphysical value 
$\delta_g < -1$.

A simple prediction of linear biasing is that the two--point correlation 
function is amplified by a factor $b^2$, while $S_3$ is inversely proportional 
to $b$. The analysis of the first redshift surveys 
revealed instead that different classes of galaxies selected in the optical
and infrared bands, while differing in the amplitude of the  two--point 
correlation function, have similar values of $S_3$ 
(\citealp{1992ApJ...398L..17G},    
\citealp{1993ApJ...417...36B},     
\citealp{1999ApJ...514..563B});      
the same also holds for galaxy clusters 
(\citealp{1995ApJ...438..507C}).     
In particular, \cite{1999ApJ...514..563B}          
analysed volume--limited samples of the Southern Sky Redshift Survey 2
(SSRS2, \citealp{1994ApJ...424L...1D})     
and found that, while the two--point correlation amplitude increases 
significantly with galaxy luminosity when $L>L_*$ 
(\citealp{1996ApJ...472..452B}),    
the value of $S_3$ does not scale with the inverse of the bias parameter $b$ 
and is independent of luminosity and scale within the errors: 
this implies that the bias is non-linear.
Similar results were obtained in the Durham/UKST and Stromlo-APM redshift 
surveys
(\citealp{2000MNRAS.317L..51H})  
and in the larger and deeper 2dF Galaxy Redshift Survey
(2dFGRS, 
\citealp{2004MNRAS.351L..44B},   
\citealp{2004MNRAS.352..828C}),  
which enabled a more detailed analysis: for example,
\cite{2004MNRAS.352.1232C} 
found evidence for a weak dependence of $S_3$ on luminosity, while 
according to 
\cite{2007MNRAS.379.1562C} 
the $S_J$ of red galaxies depends on luminosity, 
while blue galaxies do not show any dependence. 
In an analysis of the Sloan Digital Sky 
Survey (SDSS)
\cite{2006ApJ...649...48R} 
found that the values of $S_J$
are lower for late--type than for early--types galaxies.

In more recent years, deeper surveys enabled
exploring the effects of the evolution of gravitational clustering and bias, 
thus placing stronger constraints on models of galaxy formation and evolution.

\cite{2013MNRAS.tmp.2056W} 
measured the hierarchical clustering of the CFHTLS--Wide from photometric
redshifts. They found an indication that at small scales the hierarchical moments
increase with redshift, while at large scales their results are still consistent
with perturbation theory for $\Lambda CDM$ cosmology with a linear bias, but
suggest the presence of a small non-linear term.

>From the analysis of the VIMOS--VLT Deep Survey, based on spectroscopic
redshifts, \cite{2005A&A...442..801M} 
(see also \citealp{2008A&A...487....7M}) 
found that the value of $S_3$ for luminous ($M_B < -21$) galaxies is consistent 
with the local value at $z < 1$ while decreasing beyond $z\sim 1$, and
that the bias is non-linear. 

In this paper we analyse the high--order correlations and moments of the first 
release of the VIMOS Public Extragalactic Redshift Survey 
(VIPERS\footnote{http://vipers.inaf.it})
in the redshift range $0.5 < z \le 1.1$ 
as a function of luminosity and stellar mass.
We also derive an estimate of the non-linear bias.
Our analysis extends those presented in a number of 
recent works that have investigated various aspects of galaxy clustering in 
the VIPERS sample. 
Some works have focused on two--point statistics, like the standard 
galaxy-galaxy two--point correlation function to estimate redshift space 
distortions (\citealp{2013A&A...557A..54D}) 
and its evolution and dependence on galaxy properties 
(\citealp{2013A&A...557A..17M}). 
A different type of two--point statistics, the clustering ratio, has been 
introduced by 
\cite{2014A&A...563A..36B} 
and applied to VIPERS galaxies 
(\citealp{2014A&A...563A..37B}) 
to estimate the mass density parameter $\Omega_M$. 
\cite{2014A&A...570A.106M} 
have searched the VIPERS survey for galaxy voids
and characterized their properties by means of the galaxy-void 
cross-correlation. 
Bel et al. (2015, in preparation) have
proposed a method to infer the one--point galaxy 
probability function from counts in cells that 
 \cite{2014arXiv1406.6692D} 
have exploited to search for and detect deviations from linear bias;
 a result that we directly compare our results with. Finally, 
\cite{2014A&A...565A..67C} 
studied different methods for accounting for gaps in the VIPERS survey and assessing 
their impact on galaxy counts.

As cosmological parameters we have adopted $H_0 = 70$ km/s/Mpc, 
$\Omega_M = 0.25$, $\Omega_\Lambda = 0.75$, but all cosmology--dependent 
quantities are given in $H_0=100$ km/s/Mpc units associated with 
the corresponding power of $h = H_0/100$.

%
%

\section{High--order statistics}

In this section we resume the formalism and define the statistical quantities 
measured in our work. 

The volume--averaged $J$--point correlation functions are given by
\begin{equation}
 \overline{\xi}_J(V) = \frac{1}{V^J} \int _V \xi_J dV_1 ... dV_J ,
\end{equation}

where for spherical cells (used in this work) 
$\overline{\xi}_J$ is a function of the cell radius $R$ and $V = 4 \pi R^3 /3$.

The volume--averaged two--point correlation function gives the variance 
of the density contrast: 

\begin{equation}
 \sigma^2 (R) = \overline{\xi}_2(R). 
\label{eq:variance}
\end{equation}

The volume--averaged $J$--point correlation functions can be easily derived 
from the moments of
the count PDF $P(N,R)$, that is, the probability
of counting $N$ objects in a randomly chosen spherical volume of radius $R$ 
(see \citealp{1980lssu.book.....P}). 
For simplicity, in the following we omit the dependence on $R$.
At a fixed scale $R$, the centred moments of order $J$ are
\begin{equation}
\mu_J = \sum _N P(N) \left( \frac{N - \overline{N}}{\overline{N}} \right)^J ,
\end{equation}
where $\overline{N} = nV = \sum N P(N)$ is the mean number of objects in a cell
 of radius $R$.

The volume--averaged correlation functions correspond to the reduced moments 
and up to the fourth order are given by the following relations: 
\begin{align}
\overline{\xi}_2 & = \mu _2 - \frac{1}{\overline{N}} \nonumber \\
\overline{\xi}_3 & = \mu _3 - 3 \frac{\mu _2}{\overline{N}} + 
\frac{2}{\overline{N}^2} 
\nonumber \\
\overline{\xi}_4 & =  \mu _4 - 6 \frac{\mu _3}{\overline{N}} + 11 
\frac{\mu_2}{\overline{N}^2} 
- 3 \mu_2 ^2 - \frac{6}{\overline{N}^3}.
\end{align}

An alternative way to estimate the high--order correlations is through 
the factorial moments $m_k$:

\begin{equation}
 m_k = \sum _N P(N) N^{\underline{k}} ,
\end{equation}

where

\begin{equation}
 N^{\underline{k}}  \equiv  N(N-1)...(N-k+1)
\end{equation}

is the falling factorial power of order $k$ (see e.g. 
\citealp{1994cm.book.....G}). 

In fact, for a local Poisson process the moments about the origin
of a stochastic field are given by the factorial moments of $N$; 
as our variable is the number density  contrast $(N-\overline{N})/\overline{N}$,
 we have to convert the factorial moments $m_k$ into 
the moments about the mean (central moments) $\mu _k ^\prime$
through the standard relations

\begin{align}
 \mu _2 ^\prime & = m_2 - m_1^2 \nonumber \\
 \mu _3 ^\prime & = m_3 - 3 m_1 m_2 + 2 m_1^3 \nonumber \\
 \mu _4 ^\prime & = m_4 - 4 m_1 m_3 + 6 m_1^2 m_2 - 3 m_1^4. 
\end{align}

We can finally derive the volume--averaged $J$--point correlation 
functions

\begin{equation}
 \overline{\xi_J} = \frac{\mu _J ^\prime}{\overline{N}^J}  
\end{equation}

and the normalized moments $S_J$
\begin{equation}
  S_J = \frac{\overline{\xi}_J}{\overline{\xi}_2 ^{J-1}}.
\label{eq:sjclassic} 
\end{equation}

The normalized moments can also be obtained through a recursive formula 
(\citealp{1993ApJ...408...43S}, 
\citealp{2000MNRAS.313..711C}):  
\begin{equation}
 S_J = \frac{\overline{\xi}_2 N^{\underline{J}}}{N^J_c} - 
\frac{1}{J} \sum _{k=1} ^{J-1} 
\frac{J!}{(J-k)! k!} \frac{(J-k) S_{J-k} m_k}{N ^k _c},
\label{eq:sjrecursive}
\end{equation}

where 

\begin{equation}
 N_c \equiv \overline{N} \overline{\xi}_2.
\end{equation}

The values given in this paper were calculated using factorial moments.

At a fixed scale $R$, the deterministic bias parameter $b$ can be directly 
measured through the square root of the ratio of the galaxy variance
$\sigma_g ^2$ to the matter variance $\sigma_m ^2$:

\begin{equation}
 b(z) = \frac{\sigma _g (z)}{\sigma _m (z)}.
\label{eq:bias}
\end{equation}

In the case of linear biasing, the galaxy density contrast
$\delta_g$ is proportional to the matter density contrast $\delta_m$ 
by a constant factor $b$, $\delta_g = b \delta_m$: there is no dependence
on scale, and $b$ is the only parameter that completely
defines the relation between the galaxy and matter distribution.

As we have noted in the introduction, the linear biasing cannot have a general
validity. It is more general and realistic to assume a local, deterministic 
non--linear bias $b(z,\delta_m,R)$, which can be written as a Taylor expansion 
(\citealp{1993ApJ...413..447F}): 

\begin{equation}
 \delta_g = \sum _{k=0} \frac{b_k}{k!} \delta ^k , 
\end{equation}

where $b_1 \equiv b$. \cite{1993ApJ...413..447F} have shown that such a local 
bias transformation preserves the 
hierarchical properties of the underlying matter distribution in the limit
of small fluctuations (large scales).

In the case of linear bias, $b_k = 0$ for $k>1$, and the galaxy and matter 
normalized moments are then related by the following equation:

\begin{equation}
  S_{Jg} = \frac{S_{Jm}}{b^{J-1}}.
\label{eq:slinear}
\end{equation}

In general, the deviation from linear biasing is measured by
taking the second order of the expansion. 
In this case, the galaxy normalized skewness is given by the following relation:

\begin{equation}
 S_{3g} = \frac{1}{b} \left (S_{3m} + 3 \frac{b_2}{b} \right).
\label{eq:secondorderbias}
\end{equation}

%
%

\section{VIPERS survey}
\label{data-section}

\begin{table*}[t]
\caption{Definition of the samples.}     
\label{table:1}     
\centering                                     
\begin{tabular}{c c c c c c}          
\hline\hline                        
Redshift range & Limiting magnitude & $N_g$ & $\sigma_{8g}$ & $S_{3g}$  & $S_{4g}$ \\
               & $M_B (z=1.1)- 5 \log(h)$  & W1 + W4 &     & $R=8 h^{-1}$ Mpc &  $R=8 h^{-1}$ Mpc \\
\hline                                   
$0.5 \le z < 0.7$ & <-19.5 & 8670 + 6863 &$0.95 \pm 0.06 $& $1.81 \pm 0.20$ & $8.13 \pm 2.03 $ \\
$0.5 \le z < 0.7$ & <-20.0 & 6101 + 4963 &$1.00 \pm 0.06 $& $1.82 \pm 0.22$ & $8.12 \pm 2.05 $ \\
$0.5 \le z < 0.7$ & <-20.5 & 3671 + 3025 &$1.07 \pm 0.07 $& $1.83 \pm 0.27$ & $7.93 \pm 2.54 $ \\
$0.5 \le z < 0.7$ & <-21.0 & 1787 + 1478 &$1.16 \pm 0.15 $& $1.78 \pm 0.30$ & $6.29 \pm 2.96 $ \\
\hline
$0.7 \le z < 0.9$ & <-20.0 & 7455 + 5384 &$1.01 \pm 0.05 $& $1.74 \pm 0.14$ & $ 7.19 \pm 1.34 $ \\
$0.7 \le z < 0.9$ & <-20.5 & 4979 + 3475 &$1.05 \pm 0.04 $& $1.66 \pm 0.16$ & $ 6.02 \pm 1.38 $ \\
$0.7 \le z < 0.9$ & <-21.0 & 2457 + 1664 &$1.10 \pm 0.06 $& $1.59 \pm 0.22$ & $ 5.50 \pm 1.76 $ \\
\hline
$0.9 \le z < 1.1$ & <-20.5 & 2751 + 1805 &$1.12 \pm 0.07 $& $2.50 \pm 0.28$ & $14.11 \pm 3.10 $ \\
$0.9 \le z < 1.1$ & <-21.0 & 1752 + 1067 &$1.16 \pm 0.08 $& $2.54 \pm 0.38$ & $12.70 \pm 3.81 $ \\
\hline\hline                                             
Redshift range & Limiting stellar mass & $N_g$ & $\sigma_{8g}$ & $S_{3g}$ & $S_{4g}$ \\
               & $\log(M/M_{\odot} h^{-2})$  &    W1 + W4         &            &   $R=8 h^{-1}$ Mpc                &   $R=8 h^{-1}$ Mpc \\
\hline                                   
$0.5 \le z < 0.7$ & >  9.0 & 8745 + 6544 &$0.97 \pm 0.10 $& $1.88 \pm 0.15$ & $ 8.51 \pm 1.45$ \\
$0.5 \le z < 0.7$ & >  9.5 & 6091 + 4318 &$1.03 \pm 0.10 $& $1.94 \pm 0.15$ & $ 8.66 \pm 1.43$ \\
$0.5 \le z < 0.7$ & > 10.0 & 3654 + 2581 &$1.16 \pm 0.11 $& $2.02 \pm 0.16$ & $ 8.61 \pm 1.45$ \\
$0.5 \le z < 0.7$ & > 10.5 & 1292 +  713 &$1.34 \pm 0.11 $& $1.90 \pm 0.18$ & $ 6.62 \pm 1.39$ \\
\hline
$0.7 \le z < 0.9$ & >  9.5 & 6159 + 4009 &$1.09 \pm 0.08 $& $1.88 \pm 0.14$ & $ 7.59 \pm 1.37$ \\
$0.7 \le z < 0.9$ & > 10.0 & 3746 + 2428 &$1.18 \pm 0.08 $& $1.87 \pm 0.14$ & $ 7.29 \pm 1.36$ \\
$0.7 \le z < 0.9$ & > 10.5 & 1467 +  819 &$1.41 \pm 0.09 $& $2.04 \pm 0.20$ & $ 7.81 \pm 1.80$ \\
\hline
$0.9 \le z < 1.1$ & > 10.0 & 1644 + 964 &$1.23 \pm 0.08 $& $2.70 \pm 0.21$ & $ 13.28 \pm 2.43$ \\
$0.9 \le z < 1.1$ & > 10.5 &  738 + 456 &$1.43 \pm 0.09 $& $3.19 \pm 0.29$ & $ 16.18 \pm 3.88$ \\
\hline\hline

\end{tabular}
\end{table*}

The VIMOS Public Extragalactic Redshift Survey (VIPERS) is an ongoing ESO Large 
Programme aimed at determining redshifts for $\sim 10^5$ galaxies in the
redshift range 
$0.5 < z < 1.2$, to accurately and robustly measure clustering,
 the growth of structure (through redshift-space distortions) and galaxy
properties at an epoch when the Universe was about half its current
age (\citealp{2013Msngr.151...41G}; \citealp{2014A&A...566A.108G}). 
The survey is divided into two separate areas and will cover 
$\sim 24$ deg$^2$ when completed. The two areas are the so--called $W1$ and
$W4$ fields of the Canada-France-Hawaii Telescope Legacy Survey Wide 
(CFHTLS-Wide); the CFHTLS optical photometric catalogues\footnote{
Mellier, Y., Bertin, E., Hudelot, P., et al. 2008,
http://terapix.iap.fr/cplt/oldSite/Descart/CFHTLS-T0005-Release.pdf.}
constitute the parent catalogue from which VIPERS
spectroscopic targets were selected.
The VIPERS survey strategy is optimized to 
achieve a good completeness in the largest possible area 
(\citealp{2009Msngr.135...13S}). 
Galaxies are selected to a limit of $i_{AB}<22.5$, further
applying a simple and robust $gri$ colour pre-selection to 
effectively 
remove galaxies at $z<0.5$. In this way, only one pass per field is required,
 allowing us to double the galaxy sampling rate in the redshift range of 
interest with respect to a pure magnitude-limited sample ($\sim 40\%$).
The final volume of the survey will be 
$5 \times 10^{7} h^{-3}$ Mpc$^{3}$, comparable to 
that of the 2dFGRS at $z\sim0.1$.
 
VIPERS spectra are obtained using the VLT Visible Multi--Object Spectrograph
(VIMOS,
\citealp{2002Msngr.109...21L}, 
\citealp{2003SPIE.4841.1670L}) 
at moderate resolution ($R=210$), with 
the LR Red grism at $R=210$ and a wavelength coverage of 5500-9500$\rm{\AA}$.
The typical radial velocity error is $140 (1+z)$ km sec$^{-1}$.  
A discussion of the survey data reduction and the first management 
infrastructure were presented in \cite{2012PASP..124.1232G} 
and the detailed description of the survey was given by 
\cite{2014A&A...566A.108G}. 

The data set used in this and the other published papers is 
the VIPERS Public Data Release 1 (PDR-1) catalogue,
made available to the public in 2013 
(\citealp{2014A&A...562A..23G}). 
It includes about $47,000$ reliable spectroscopic redshifts of galaxies and 
active galactic nuclei (AGNs).
We here only selected galaxies with reliable redshift,
that is, with spectroscopic quality flags $2, 3, 4$, or $9$ 
(see \citealp{2014A&A...562A..23G} for the definition).

To avoid regions dominated by large gaps,
we here selected a subset of the total area covered by VIPERS: 
our limits are 
$02^h 01^m 00^s \le RA \le 02^h 34^m 50^s$, $-5.08^o \le DEC \le -4.17^o$ 
(7.67 square degrees) in $W1$ and  
$22^h 01^m 12^s \le RA \le 22^h 18^m 00^s$, $0.865^o \le DEC \le 2.20^o$
(5.60 square degrees) in $W4$. 
 
We defined volume--limited subsamples with different absolute magnitude 
and stellar mass limits, following the same criteria as in 
\cite{2013A&A...557A..17M}. 
The choice of these particular samples is discussed in detail in that paper;
here we recall their main properties.

The rest--frame B--band absolute magnitude and the stellar mass were estimated 
through the HYPERZMASS program 
(\citealp{2000A&A...363..476B}, 
\citealp{2010A&A...524A..76B}),  
which applies a spectral energy distribution (SED) fitting technique.  
To take into account luminosity evolution, we fixed as a reference limit
the luminosity at our maximum redshift ($z = 1.1$) and assumed an evolution
$M(z) = M(0) - z$ (see \citealp{2009A&A...505..463M} 
and also \citealp{2005A&A...439..863I}, 
\citealp{2009A&A...508.1217Z}). 

We did not correct the mass 
limit of the stellar-mass-limited subsamples; this limit was therefore kept fixed within each redshift bin because the evolution of $M_*$ is negligible in our redshift range 
(\citealp{2007A&A...474..443P}, 
\citealp{2010A&A...523A..13P}, 
\citealp{2013A&A...558A..23D}) 


   \begin{figure*}[t]
   \centering
      \includegraphics[width=17cm]{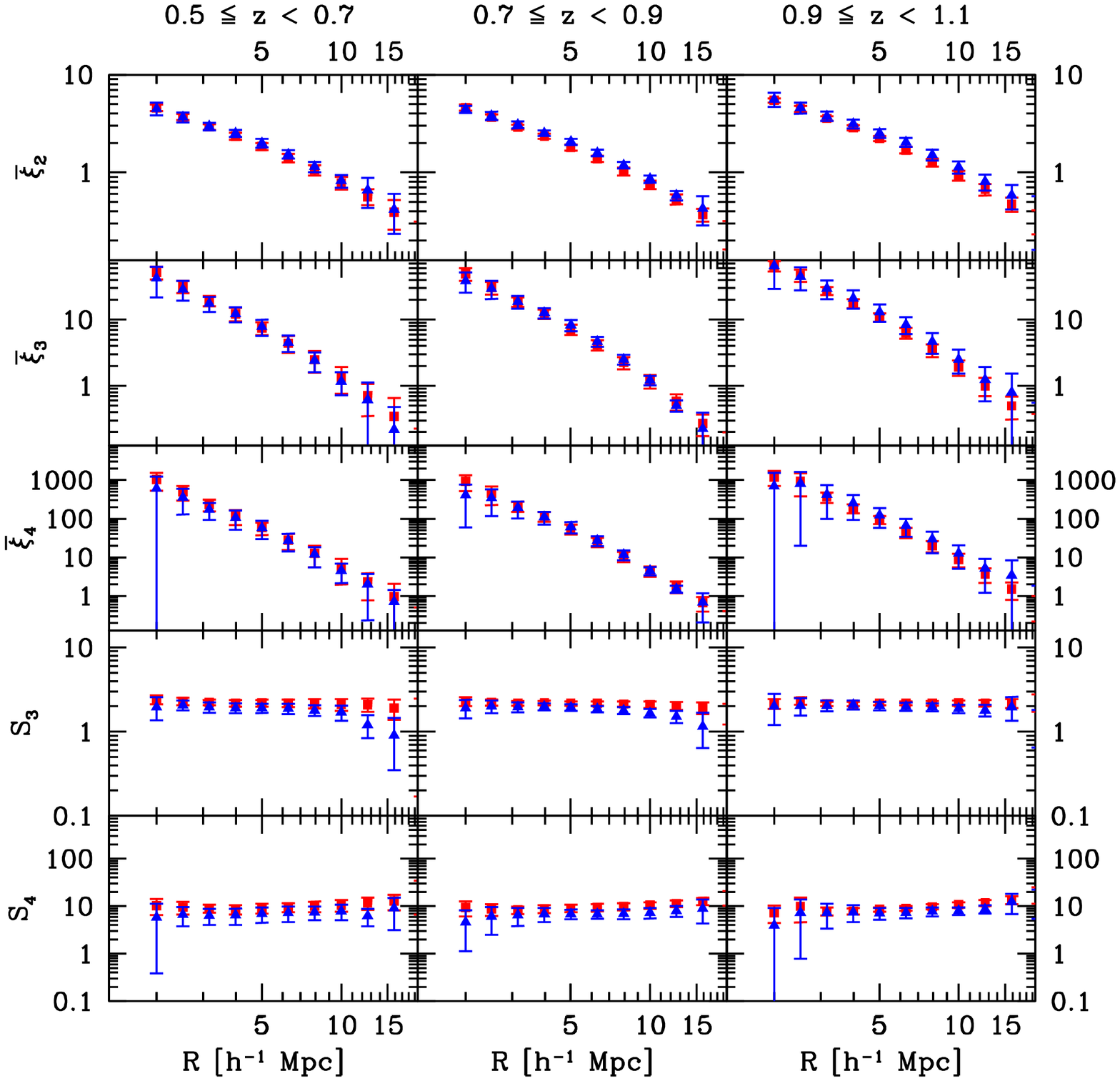}
      \caption{Comparison between mock catalogues with a sampling
rate of 100\% 
 and without gaps (red triangles), and with a sampling rate and gaps 
as in VIPERS (blue triangles). The subsamples are limited at 
$M_B(z=1.1)-5\log(h) \le -20.5$.
>From top to bottom: volume--averaged two-, three-, and four--point correlation 
functions, normalized skewness $S_3$ and kurtosis $S_4$ in redshift space.
First column: $0.5 \le z < 0.7$; second column: $0.7 \le z < 0.9$; 
third column: $0.9 \le z < 1.1$.}
         \label{fig:mocks}
   \end{figure*}

The respective numbers of galaxies for the different subsamples are given in 
Table \ref{table:1}. We note that these numbers are slightly different from 
those in \cite{2013A&A...557A..17M} because we applied more stringent 
angular limits to avoid regions nearby prominent gaps that might 
affect the counts in spherical cells (while the direct estimate of the
two--point correlation function through counts of galaxy pairs can be easily 
corrected for by using a random catalogue with the same survey geometry).

%
%

\section{Analysis of mock catalogues}


We used mock catalogues derived from cosmological simulations to 
estimate not only the statistical errors and the uncertainty related to cosmic 
variance, but also the systematic errors that are due to the inhomogeneous 
spectroscopic completeness and the specific geometry of the two fields.
A detailed description of the way these mocks were built was given by
\cite{2013A&A...557A..54D}. 

We analysed a set of 26 independent mock catalogues based on the dark 
matter halo catalogue of the MultiDark simulation 
(\citealp{2012MNRAS.423.3018P}), 
which assumes a flat
$\Lambda CDM$ cosmology with ($\Omega_M$, $\Omega_\Lambda$, $\Omega_b$, 
$h$, $n$, $\sigma_{8m}$) = (0.27, 0.73, 0.0469, 0.7, 0.95, 0.82). 
This catalogue was populated with galaxies using halo occupation distribution 
prescriptions, as described in \cite{2013A&A...557A..54D}. 
In particular, the original halo catalogue was repopulated with halos below 
the resolution limit with the new technique of 
\cite{2013MNRAS.435..743D}, 
which enables reproducing the range in stellar mass and luminosity 
probed by VIPERS data. 
For luminosity--limited subsamples, galaxy luminosities were calibrated using
VIPERS data, while for stellar mass--limited subsamples 
masses were assigned to galaxies using 
the stellar-to-halo mass relation (SHMR) of 
\cite{2013MNRAS.428.3121M}. 
 From the parent mock catalogues, a set of spectroscopic catalogues was derived by 
applying the same angular, photometric, and spectroscopic selection functions 
as were applied to the real data.
For a more detailed and complete description of the mock catalogues see 
\cite{2013A&A...557A..54D}. 


   \begin{figure}[ht]
   \centering
      \includegraphics[width=9.5cm]{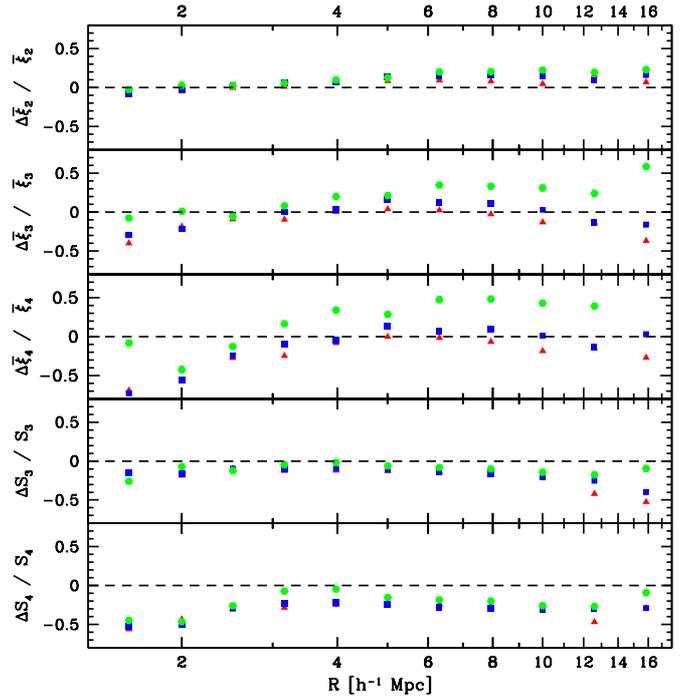}
      \caption{Fractional difference of the average
$\overline{\xi_2}$, $\overline{\xi_3}$, $\overline{\xi_3}$, $S_3$ and $S_4$ 
(from top to bottom) for the same set of mock catalogues as defined in Fig. 1,
i.e. with 100\% sampling rate and without gaps, and with sampling rate and 
gaps as in VIPERS. The subsamples are limited at 
$M_B(z=1.1)-5\log(h) \le -20.5$. 
Red triangles: $0.5 \le z < 0.7$;
blue squares: $0.7 \le z < 0.9$; green hexagons: $0.9 \le z < 1.1$.}
         \label{fig:diffmocks}
   \end{figure}


>From the mock spectroscopic catalogues we derived volume--limited subsamples 
with cuts in blue absolute magnitude and stellar mass corresponding to the 
observed ones.
First of all, these mocks were used to test the effect of the gaps in the 
survey.
As VIMOS is made of four quadrants $7\times8$ separated by 2 arcmin,
characteristic cross--shaped gaps are left in the survey; a further gap is 
present between the rows of pointings at different declination; finally,
there are a few missing quadrants due to failed pointings. Cells whose 
projection on the sky includes a gap can potentially miss some galaxies, 
which affects final counts.

These gaps might be avoided by conservatively only counting galaxies in the cells that are completely included in one single quadrant, but in this way, only small scales would be sampled (the exact value
 obviously depends on the cell distance but it is generally lower than 
$R \sim 5 h^{-1}$ Mpc). 
Alternatively, the counts in each cell might be associated with the effective 
volume of the cell, subtracting the volume falling into the gaps; but this less 
drastic choice, which would slightly alter the shape of the cells, 
would still limit the range of the sampled scales.

Another option would be filling the gaps.
\cite{2014A&A...565A..67C} 
applied two algorithms that use the photometric redshift information and 
assign redshifts to galaxies based upon the spectroscopic redshifts of the 
nearest neighbours. 
In this way, it is also possible to take into account the varying 
completeness from field to field. Tests on mocks have shown that these 
algorithms are successful in reconstructing the lowest and highest density 
environments at a scale of 5 $h^{-1}$ Mpc, but not in recovering the count 
PDF and its moments due to systematic biases.

We therefore here adopted another solution. 
The tests on mocks have shown that when cells are not 
allowed to cross the gaps by more than 40\% of their volume, 
the non--observed regions and the varying sampling rate can be approximated 
by a random Poisson sampling, and the original count PDF can be recovered with 
good precision (Bel et al. 2015, in preparation).
This means that to obtain good estimates of the quantities we
discuss here ($J$--point correlations and normalized moments), 
which depend  on the density contrast $\Delta \rho / \rho$, 
it is sufficient to implement the restriction 
on the volume of the cells falling into the gaps. 

In our analysis, we conservatively only considered spherical cells 
for which no more than 30\% of the volume falls in a gap.
Moreover, to improve the statistics, 
we combined the counts of the $W1$ and $W4$ fields.

In Fig. \ref{fig:mocks} we show the results obtained from the
analysis of mock subsamples limited at $M_B(z=1.1)-5\log(h) \le -20.5$ in
the three redshift bins [0.5,0.7], [0.7,0.9], [0.9,1.1]. 
We compare the ideal case with 100\% completeness and 
no gaps to the more realistic case with gaps and the same spectroscopic 
incompleteness as in our observed catalogue, that is, including the effects of the 
target sampling rate, $TSR(Q)$, and the spectroscopic sampling rate, $SSR(Q)$, 
where $Q$ indicates the quadrant dependence. 

Two other selection effects were
not taken into account: the colour sampling rate, $CSR(z)$, 
and the small-scale bias due to the constraints in the spectroscopic 
target selection (slits cannot overlap).
The first effect depends on redshift but it is weak in our redshift range 
(see Fig. 5 of \citealp{2014A&A...566A.108G}), 
while the second effect is negligible because
the angular radii of our cells are generally larger than the 
size of one quadrant.

We note that other sources of systematic errors,
as discussed by \cite{1999ApJ...519..622H}, 
are the integral constraint bias, affecting the  $J$--point correlation 
functions,
and the ratio bias, affecting the estimate of $S_J$. Given the large size of
our volumes, such systematic effects are weaker than the other 
errors, however, and can be neglected.

Figure \ref{fig:mocks} shows that the original values are recovered with
good precision (within 1$\sigma$ error), particularly in the scale range 
between $4$ and 10 $h^{-1}$ Mpc.

A more detailed analysis of the differences is possible with 
Fig. \ref{fig:diffmocks}, which gives the fractional difference for 
$\overline{\xi_2}$, $\overline{\xi_3}$,  $\overline{\xi_4}$, $S_3$ , and $S_4$ as 
a function of scale for the same mock subsamples as in Fig. \ref{fig:mocks}:
it shows that in most cases we can retrieve the $J$--point correlation 
functions and $S_J$ with only a small systematic difference. 
In the first redshift bin ($0.5 \le z < 0.7$) at a radius $R = 8 h^{-1}$ Mpc, 
$\overline{\xi_2}$ is overestimated by 
~8\%, while $\overline{\xi_3}$ is underestimated by ~3\% and
$\overline{\xi_4}$ by ~6\%: this translates 
into an underestimate of $S_3$ by ~16\% and of $S_4$ by ~26\%.
We have similar values in the second redshift bin ($0.7 \le z < 0.9$).
In the last redshift bin ($0.9 \le z < 1.1$) 
the $J$--point correlation functions
show the largest difference, increasing with order $J$: 
but  these deviations at different orders are correlated,
so that  finally the values of $S_3$ at 8$h^{-1}$ Mpc is
underestimated by only 10\% and of $S_4$ by 20\%, which is comparable
to what is found for the other two redshift bins.
The cause of the larger deviations in the last redshift bin is the lower 
density of the subsample; we take
these systematics into account in the discussion of our results.


   \begin{figure*}[t]
     \centering
       \resizebox{\hsize}{!}{\includegraphics{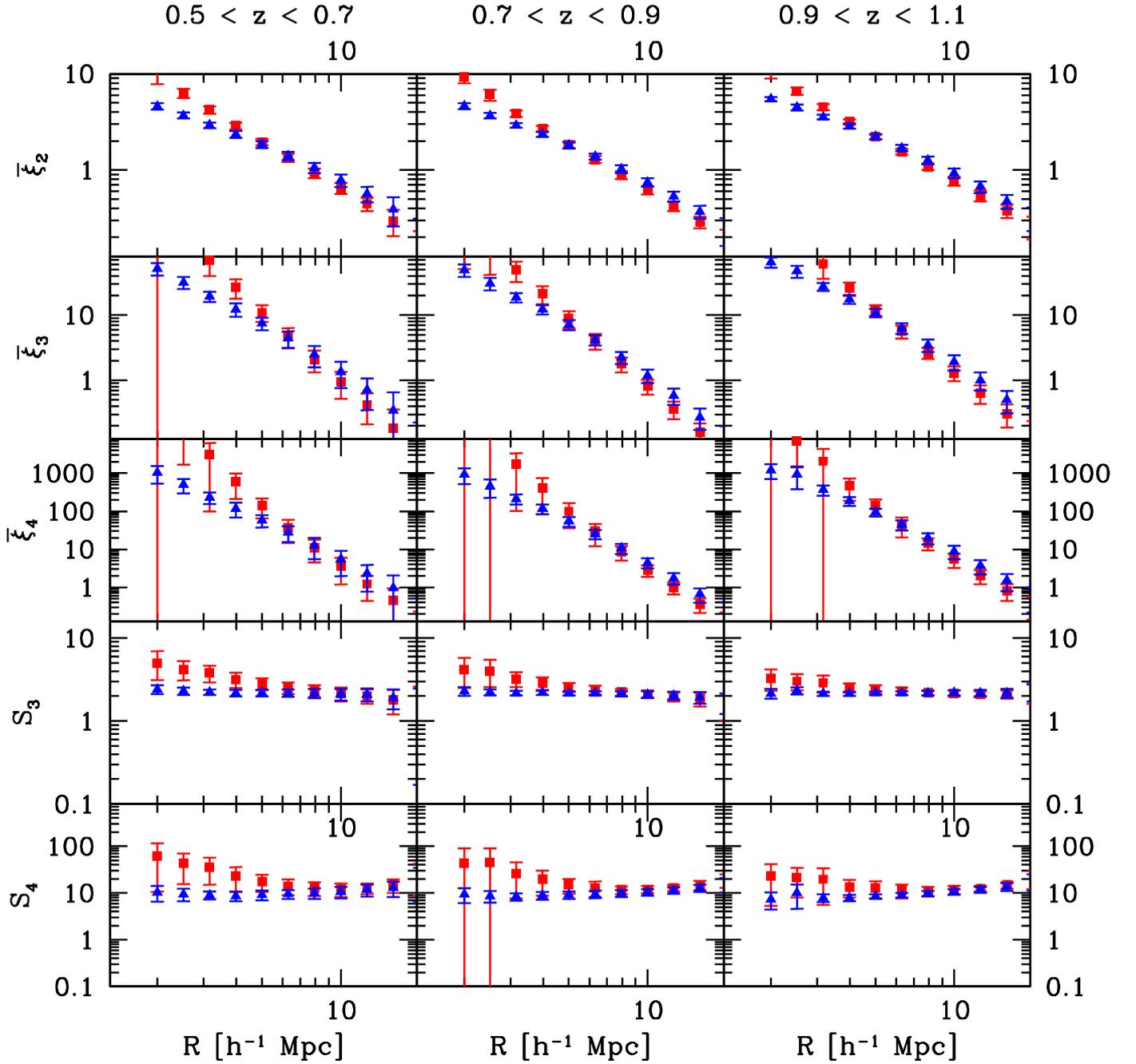}}
      \caption{Comparison between mock catalogues with 100\% sampling rate
and without gaps in real space (red squares) and redshift space 
(blue triangles). The subsamples are limited at $M_B(z=1.1)-5\log(h) \le -20.5$.
First column: $0.5 \le z < 0.7$; second column: $0.7 \le z < 0.9$; 
third column: $0.9 \le z < 1.1$.}
         \label{fig:realspace}
   \end{figure*}


It is interesting to point out that we find 
values between 1.8 and 2.1 for $S_3$ and between 8 and 10 for $S_4$ for mocks;
as an example, the analysis of the mock subsamples 
limited at $M_B(z=1.1)-5\log(h) \le -20.5$ in the redshift bin [0.7,0.9]
gives $S_3 \sim 2.13 \pm 0.16 $ and $S_4 \sim 9.8 \pm 1.6$ at 
$R = 8 h^{-1}$ Mpc. 
$S_3$ and $S_4$ show no significant redshift evolution, and their values are 
also comparable within the errors to the value measured in local redshift 
surveys for galaxies in a similar luminosity range. 

Because we know both the cosmological and the 
``observed'' redshift for galaxies in the mock samples, including the peculiar velocity and measurement error, we can estimate the conversion factor from redshift to 
real space from the mock samples. We need this factor to compare our results with second--order 
perturbation theory predictions. 
Figure \ref{fig:realspace} shows the difference between the estimates in real
and redshift space for the subsamples limited at 
$M_B(z=1.1)-5\log(h) \le -20.5$ in the three redshift ranges. 
The redshift space correlation functions show the expected loss of power at
small scales and the reverse trend at large scales. 
The estimate of the volume--averaged two--point correlation function 
in redshift space is flatter than the corresponding estimate in real space;
the difference becomes significant on scales smaller than 
$\sim 4 h^{-1}$ Mpc.
While the real space values of $S_3$ and $S_4$ increase at smaller scales, 
the increase is suppressed in redshift space; the 
difference becomes small beyond $\sim 4 h^{-1}$ Mpc.
However, at small scales we have large errors due to the 
small number of objects in the cells. 
For these reasons we focus our analysis on the 4--10 $h^{-1}$ Mpc range,
and particularly at 8 $h^{-1}$ Mpc, 
where we expect to be in the quasi--linear regime and 
predictions of second--order perturbation theory should hold.

We recall here another bias affecting mass--selected galaxy samples, 
which has been discussed and tested with mock catalogues by 
\cite{2013A&A...557A..17M}.
The lowest stellar mass subsamples suffer from incompleteness because VIPERS is magnitude limited ($i_{AB} < 22.5$); as a
consequence, we can miss high mass--to--light ratio galaxies. 
>From the analysis of mocks, \cite{2013A&A...557A..17M} found that
these galaxies are faint and red and that the 
clustering amplitude can be suppressed up to 50\% on scales below 
1 $h^{-1}$ Mpc. However, as discussed by 
\cite{2013A&A...557A..17M}, 
the abundance of red and faint galaxies is overpredicted by the semi--analytic 
model used for the tests, and the clustering of red galaxies appears to be
overestimated with respect to real data
(\citealp{2011A&A...525A.125D}, 
 \citealp{2012A&A...548A.108C}), 
so that the amplitude of the effect might be overestimated.
As we have previously noted, we did not analyse 
small scales and did not correct for stellar mass incompleteness.

%
%

\section{Results}

%
%

\subsection{Volume--averaged correlation functions}

In this section we present the results of our statistical analysis on the 
combined $W1$ and $W4$ samples.

%
%

   \begin{figure*}
   \centering
      \includegraphics[width=17cm]{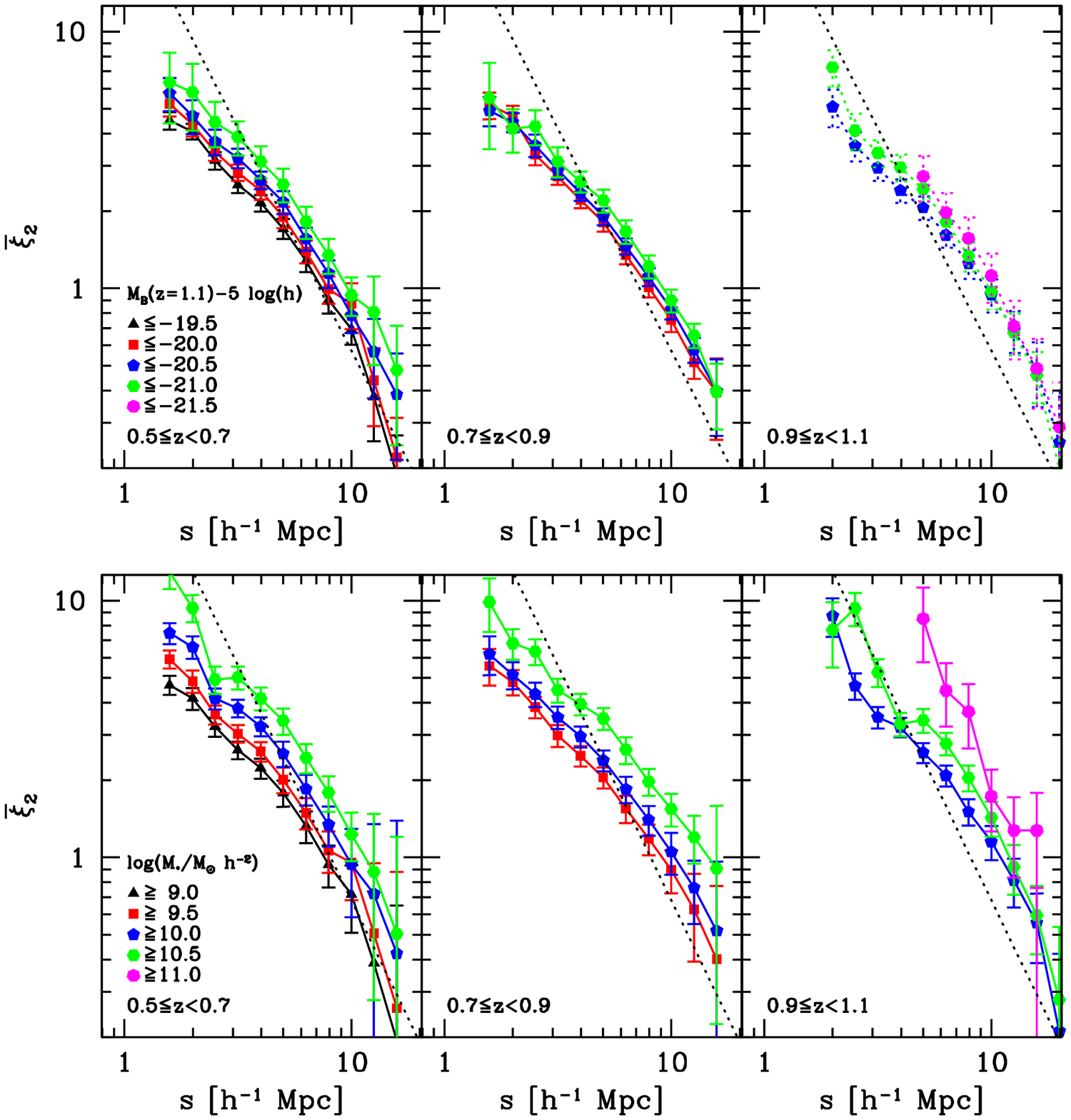}
      \caption{Volume--averaged two--point correlation functions 
$\overline{\xi_2}$ as a function 
of the $B$ absolute luminosity (upper panels) and stellar mass (lower panels).
The limits of the subsamples in absolute magnitude and stellar mass are shown 
 in the left upper and 
lower panels.
Black triangles: $M_B(z=1.1)-5\log(h)\le -19.5$ 
($\log(M_*/M_\odot h^{-2} \ge 9.0 M_\odot$); 
red squares: $M_B(z=1.1)-5\log(h) \le -20.0$ 
($\log(M_*/M_\odot h^{-2} \ge 9.5 M_\odot$); 
blue pentagons: $M_B(z=1.1)-5\log(h) \le -20.5$ 
($\log(M_*/M_\odot h^{-2} \ge 10.0 M_\odot$); 
green hexagons: $M_B(z=1.1)-5\log(h) \le -21.0$ 
($\log(M_*/M_\odot h^{-2} \ge 10.5 M_\odot$); 
magenta heptagons: $M_B(z=1.1)-5\log(h) \le -21.5$ 
($\log(M_*/M_\odot h^{-2} \ge 11.0 M_\odot$). 
Dotted lines: real--space 
$\overline{\xi_2}$ for the $M_B(z=1.1)-5\log(h) \le -20.5$ 
($\log(M_*/M_\odot h^{-2} \ge 10.0 M_\odot$) subsamples in the redshift bin 
[0.5,0.7], predicted from the power--law fit of $\xi_2$ in 
\cite{2013A&A...557A..17M}.}
         \label{fig:xi2}
   \end{figure*}

Figure \ref{fig:xi2} shows the volume--averaged two--point correlation 
function obtained from counts in cells for luminosity- and stellar 
mass--limited subsamples in the three different redshift bins.

In the same figure, as a reference for comparing the results in the 
different redshift bins, we plot the expected {\em \textup{real space}}
power--law $\overline{\xi_2}$ in the redshift bin [0.5,0.7] 
for the  $M_B(z=1.1)-5\log(h) \le -20.5$ 
subsamples (top panels) and  $M_* \ge 10.0 M_\odot$ (bottom panels), 
derived from the $\xi_2$ estimate of \cite{2013A&A...557A..17M}; we
converted their two--point correlation function to the volume--averaged 
correlation function through the formula (\citealp{1976A&A....53..131P}):

\begin{equation}
 \overline{\xi}_2 = \frac{72}{2^\gamma (3-\gamma)(4-\gamma)(6-\gamma)} \xi_2 .
\end{equation}

The line shows the effects of redshift space
distortions, which lower the value of $\overline{\xi}_2$ on small scales and
increase it on large scales.

It is clear that the amplitude of 
$\overline{\xi}_2$ increases with both luminosity and stellar mass at all redshifts.
$\overline{\xi}_2$ appears to have a stronger dependence on stellar
mass than on luminosity, in agreement with the results of 
\cite{2013A&A...557A..17M}: see their Fig.3 for the redshift
space two--point correlation functions. 

There are some fluctuations: for example, the dependence on luminosity 
appears to be sligthly weaker in the intermediate and distant redshift bins.
However, these variations are consistent when taking
into account statistical errors and sample variance, which are included in 
error bars.
We conclude that the dependence of
the two--point correlation function on luminosity and stellar mass 
does not evolve significantly up to $z \sim 1$.

%
%

   \begin{figure*}
   \centering
      \includegraphics[width=17cm]{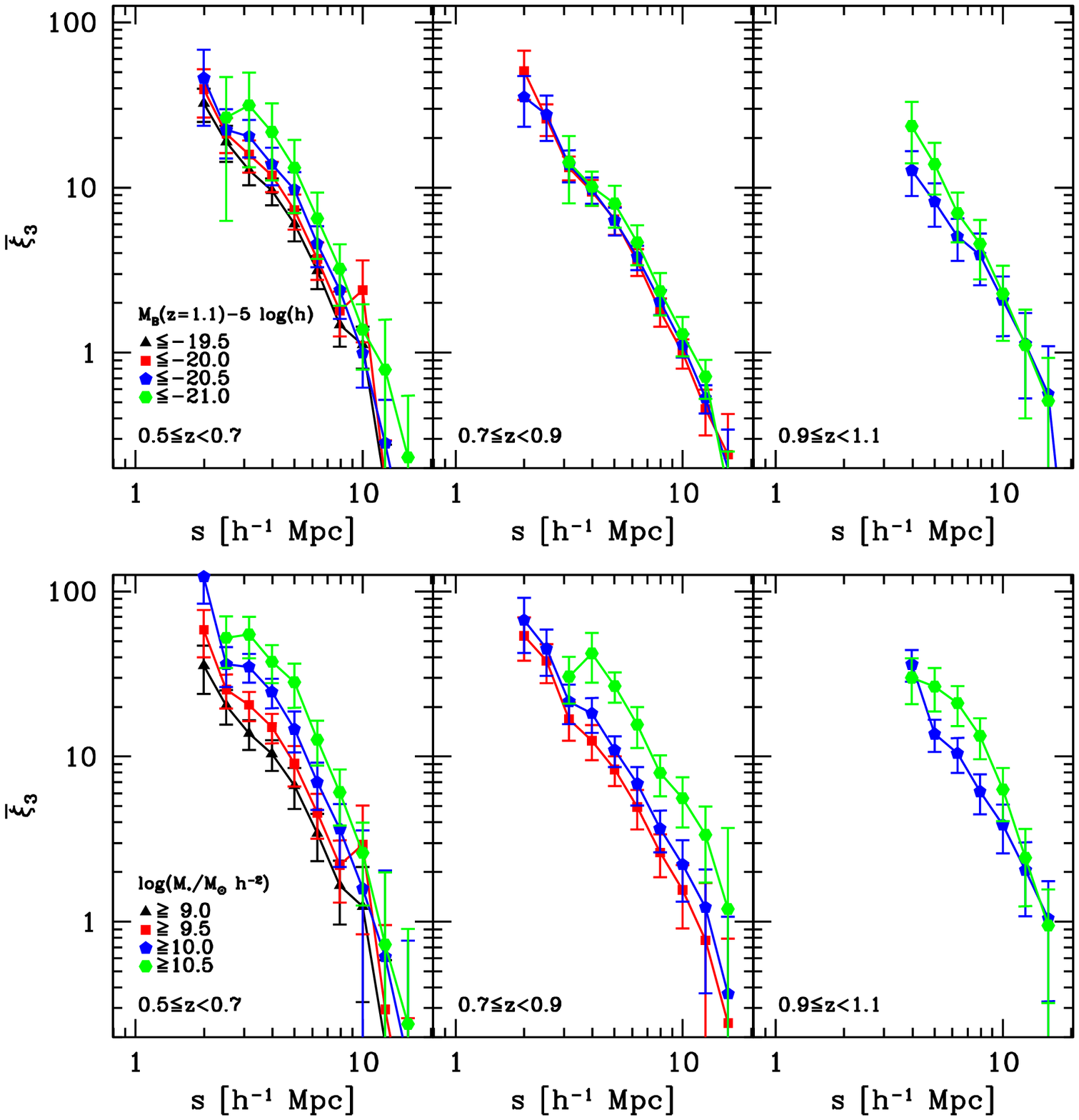}
      \caption{Volume--averaged three-point correlation functions as a function
of the $B$ absolute luminosity (upper panels) and stellar mass 
(lower panels). 
The limits of the subsamples in absolute magnitude and stellar mass are shown in the left upper and lower panels 
(symbols and colours are the same as in Fig. \ref{fig:xi2}).}
         \label{fig:xi3}
   \end{figure*}

%
%

   \begin{figure*}
   \centering
      \includegraphics[width=17cm]{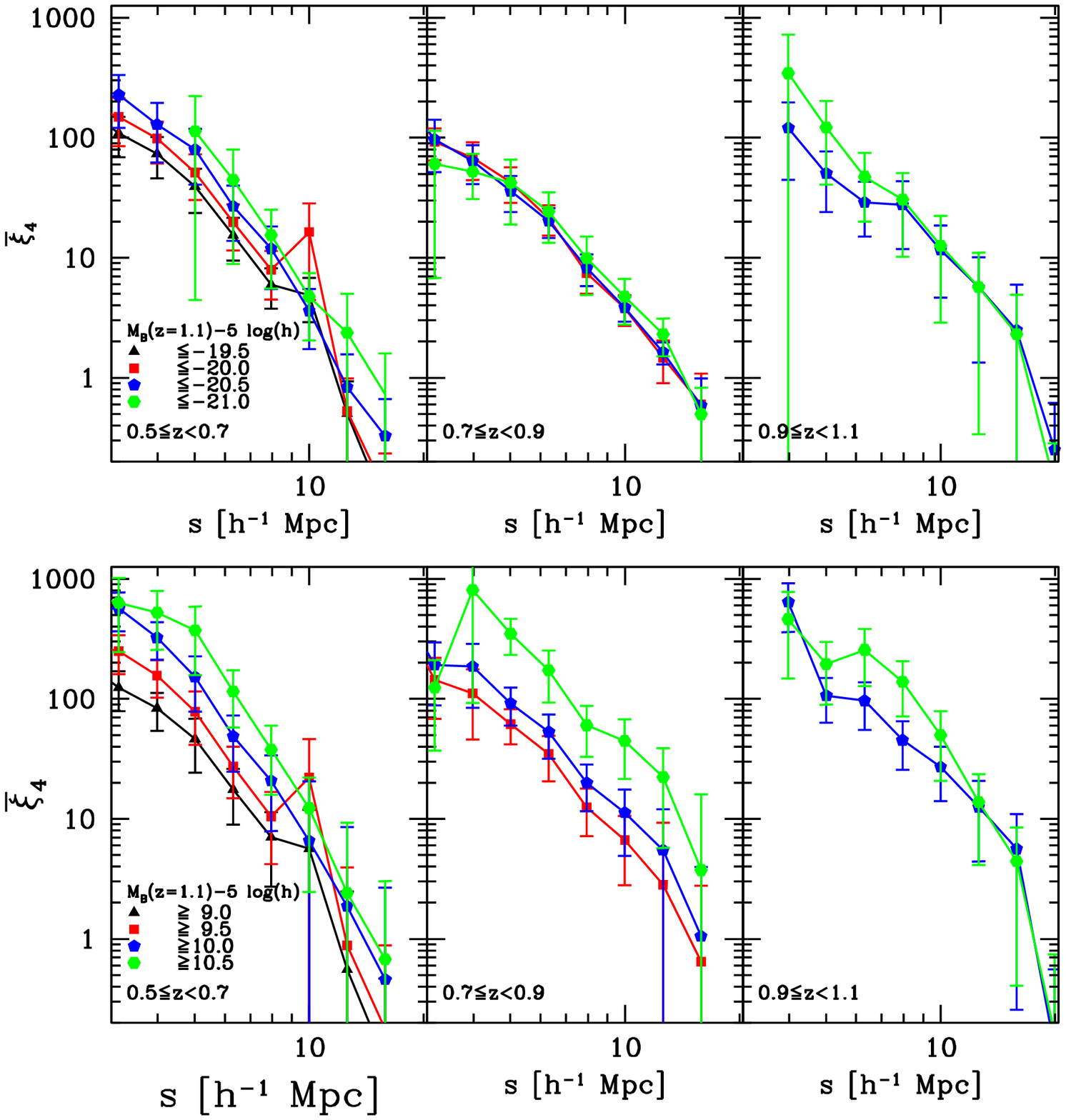}
      \caption{Volume--averaged four--point correlation functions as a 
function of the $B$ absolute luminosity (upper panels) and stellar mass 
(lower panels). 
The limits of the subsamples in absolute magnitude and stellar mass are shown in the left upper and lower panels 
(symbols and colours are the same as in Fig. \ref{fig:xi2}).}
         \label{fig:xi4}
   \end{figure*}

In Figs. \ref{fig:xi3} and \ref{fig:xi4} we show 
the volume--averaged three- and four--point correlation functions. 
Their behaviour reflects the two--point correlation functions,
showing a stronger dependence of
the correlation amplitude on stellar mass than on luminosity.

%
%

   \begin{figure*}
   \centering
      \includegraphics[width=17cm]{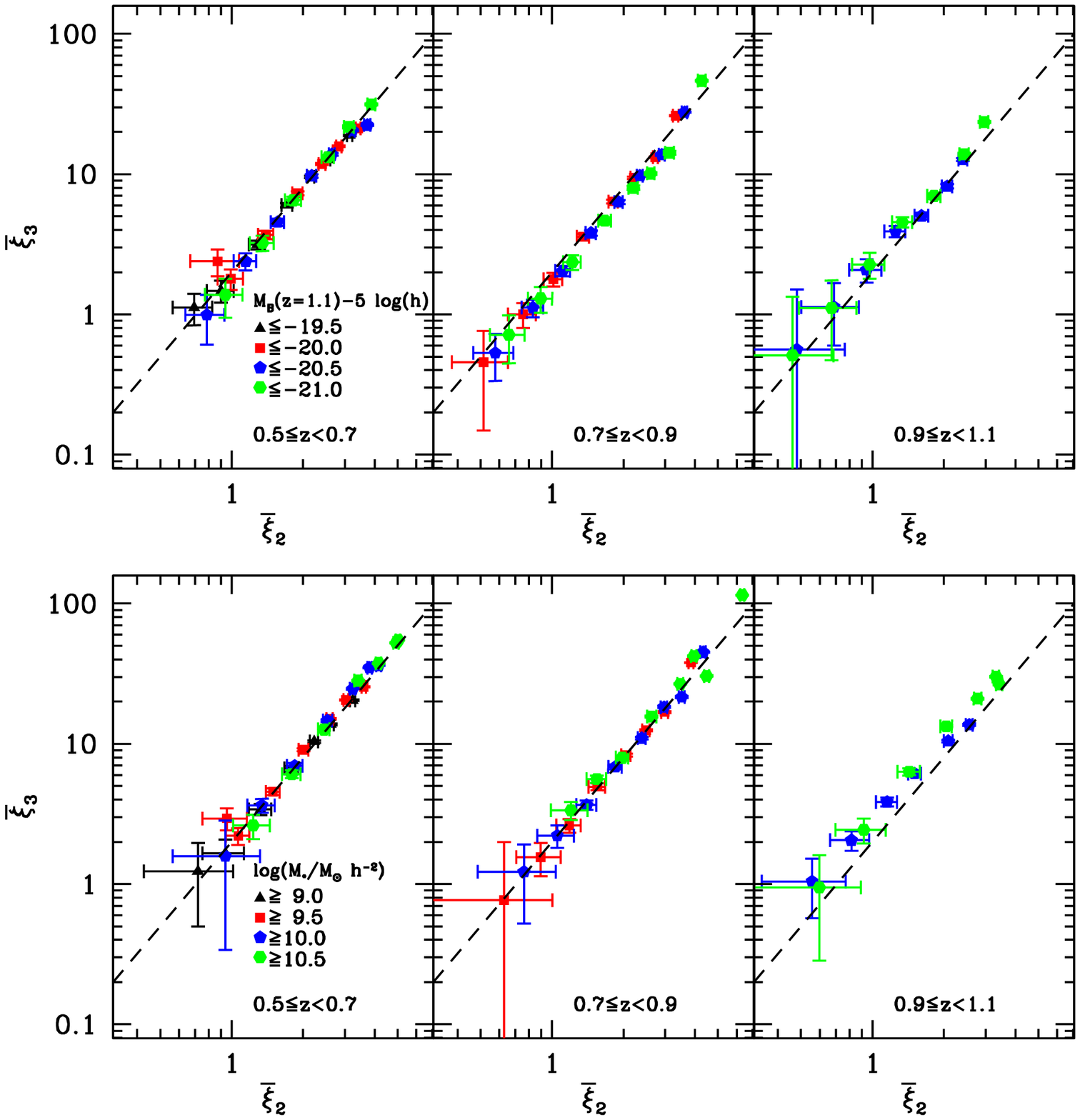}
      \caption{Scaling relation of the volume--averaged two-- and three-point 
correlations as a function of the
 $B$ absolute luminosity (upper panels) and stellar mass (lower panels). 
The limits of the subsamples in absolute magnitude and stellar mass are shown 
in the left upper and lower panels 
(symbols and colours are the same as in Fig. \ref{fig:xi2}).
The dashed line represents the scaling relation 
$\log_{10}(\overline{\xi}_3) = 2 \log_{10}(\overline{\xi}_2) + \log_{10}(2)$.} 
         \label{fig:x2x3}
   \end{figure*}

%
%

   \begin{figure*}
   \centering
      \includegraphics[width=17cm]{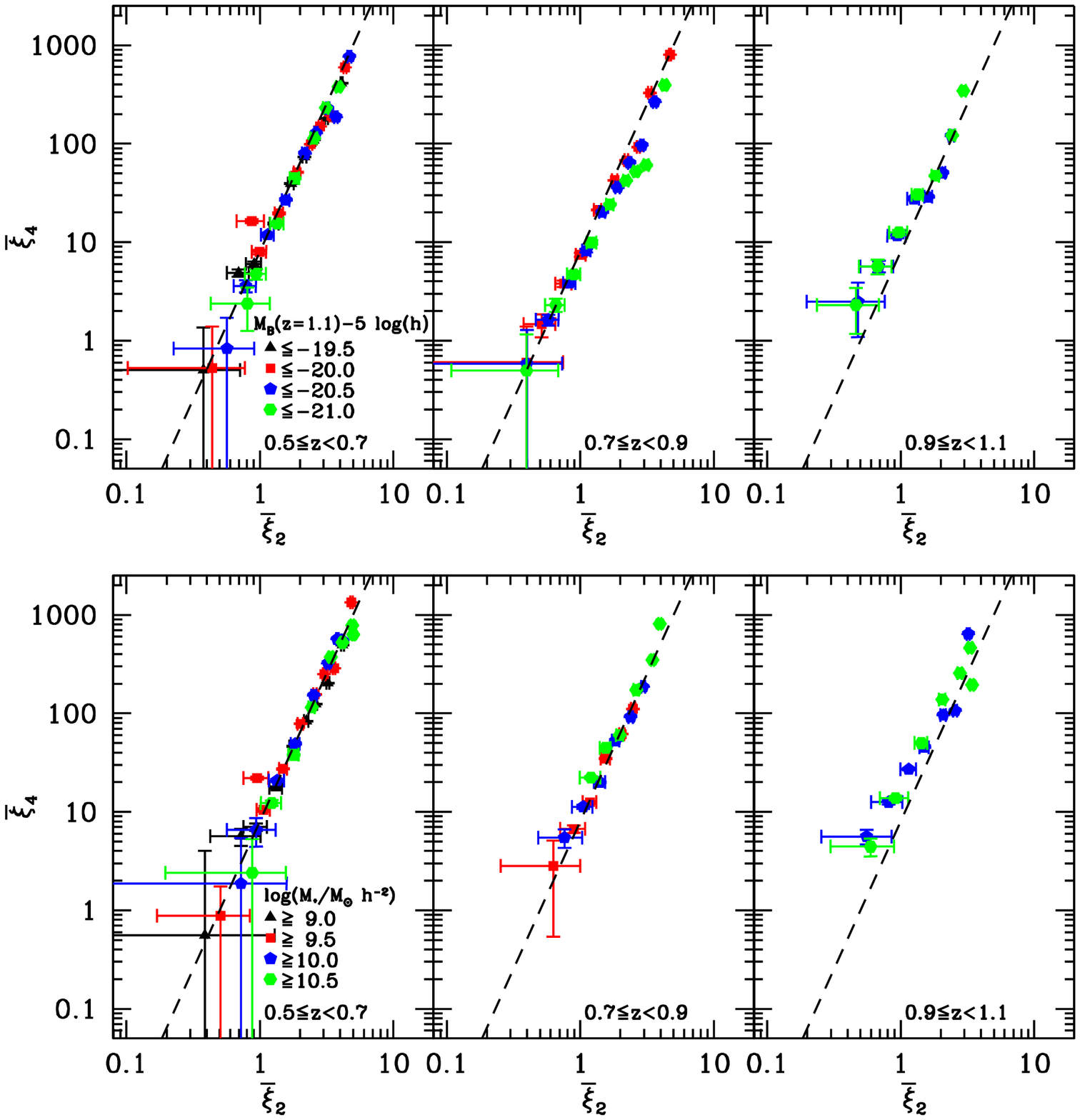}
      \caption{Scaling relation of the volume--averaged two and four--point 
correlations as a function of the
 $B$ absolute luminosity (upper panels) and stellar mass (lower 
panels). 
The limits of the subsamples in absolute magnitude and stellar mass are shown 
 in the left upper and lower panels 
(symbols and colours are the same as in Fig. \ref{fig:xi2}).
The dashed line represents the scaling relation 
$\log_{10}(\overline{\xi}_3) = 3 \log_{10}(\overline{\xi}_2) + \log_{10}(8)$.} 
         \label{fig:x2x4}
   \end{figure*}

The specific signature of the hierarchical scaling is the power--law relation
between high--order correlation functions (Eq. \ref{eq:sjclassic}).
In Figs. \ref{fig:x2x3} and \ref{fig:x2x4} we show 
the three- and four--point 
volume--averaged correlation functions as a function of the two--point 
volume--averaged correlation functions. 
The data clearly follow the hierarchical scaling relations 
$\overline{\xi}_3 \propto
 \overline{\xi}_2 ^2$ and $\overline{\xi}_4 \propto \overline{\xi}_2 ^3$.
These relations appear to hold at all luminosities and masses in the the 
first two redshift bins, but some systematic differences appear in the last 
redshift bin, particularly for the stellar--mass limited subsamples, 
where points are systematically higher than the reference scaling law, but in this case the values are also consistent with the same scaling relation 
observed at lower redshifts.

As we have previously discussed, the existence of these scaling relations 
has been verified in the local Universe: they are expected for the matter 
distribution in the quasi--linear regime, as a consequence of gravitational 
clustering. In this case, it is natural that they do not evolve with redshift:
however, it is not an obvious result to observe the same hierarchical 
behaviour for the galaxy distribution at all redshifts, given the evolution
of bias.

%
%

\subsection{Skewness and kurtosis}

>From the counts in cells we derived the rms $\sigma$ (Eq.
 \ref{eq:variance}), the normalized 
skewness $S_3$ and kurtosis $S_4$ (Eq. \ref{eq:sjrecursive}) 
for the different VIPERS subsamples. Their 
values at $R = 8 h^{-1}$ Mpc are given
in Cols. (4), (5), and (6) of Table 
\ref{table:1}. The $R=8 h^{-1}$ Mpc reference radius is nearly optimal 
because it is large enough to enter into the quasi--linear regime, and at the 
same time it is in the scale range for which we have a good sampling.

\begin{figure*}
   \centering
      \includegraphics[width=17cm]{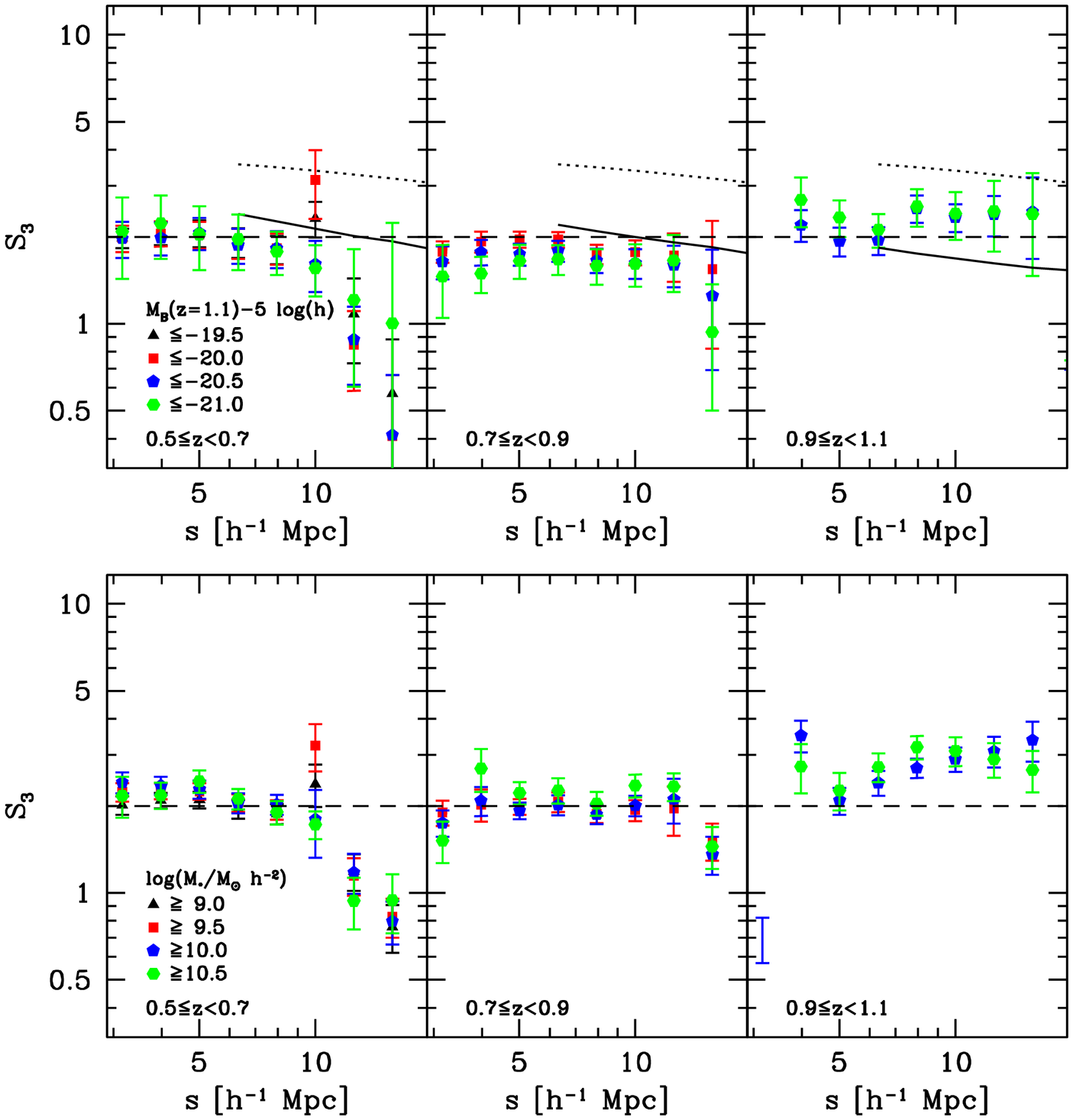}
      \caption{Normalized skewness $S_3$ as a function of the $B$ absolute 
luminosity (upper panels) and stellar mass (lower panels). 
The limits of the subsamples in absolute magnitude and stellar mass are shown 
 in the left upper and lower panels 
(symbols and colours are the same as in Fig. \ref{fig:xi2}.
We note that black points corresponding to galaxies with $M< -19.50$ or 
$M_* \ge 9.0 M_\odot$ are only plotted 
for the first redshift bin ([0.5, 0.7]), but most of them are not visible as 
they lie below the points
of the other samples. The dashed line corresponds to $S_3 = 2$.
In the top panels, the dotted line is the prediction from second--order 
perturbation theory for the matter distribution in real space; 
the solid line is the prediction of $S_3$ for galaxies with 
$M_B(z=1.1)-5\log(h) \le -20.5$ (to be compared to blue pentagons): it was
obtained from the matter values, converted to redshift space
and divided by the corresponding linear bias factor.
}
         \label{fig:s3}
   \end{figure*}

\begin{figure*}
   \centering
      \includegraphics[width=17cm]{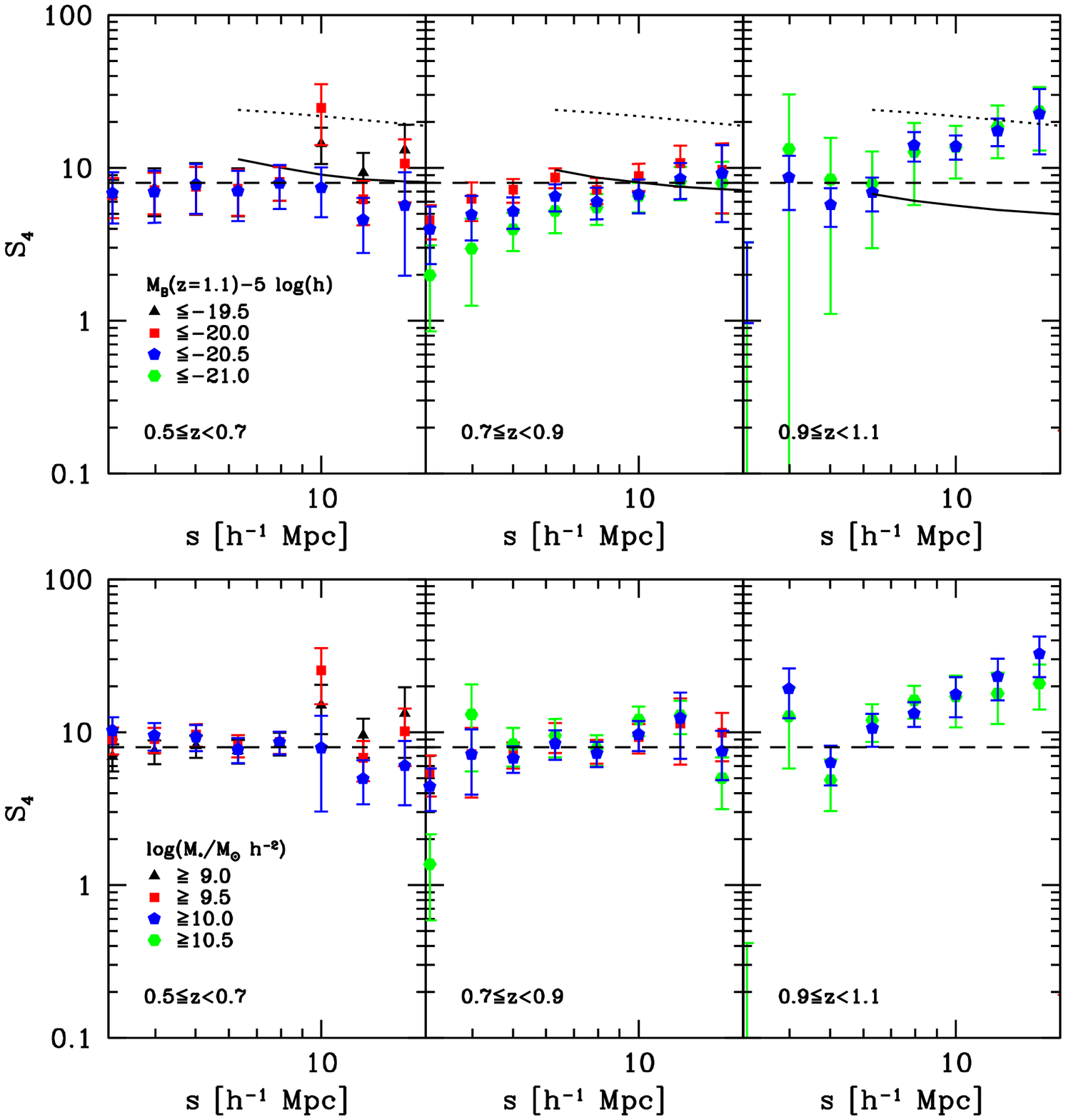}
      \caption{Normalized skewness $S_4$ as a function of the $B$ absolute 
luminosity (upper panels) 
and stellar mass (lower panels). 
The limits of the subsamples in absolute magnitude and stellar mass are shown 
 in the left upper and lower panels 
(symbols and colours are the same as in Fig. \ref{fig:xi2}).
We note that black points corresponding to galaxies with $M< -19.50$ or 
$M_* \ge 9.0 M_\odot$ are only plotted 
for the first redshift bin ([0.5, 0.7]), but most of them are not visible as 
they lie below the points
of the other samples. The dashed line corresponds to $S_4 = 8$.
In the top panels, the dotted line is the prediction from second--order 
perturbation theory for the matter distribution in real space; 
the solid line is the prediction of $S_4$ for galaxies with 
$M_B(z=1.1)-5\log(h) \le -20.5$ (to be compared to blue pentagons): it was
obtained from the matter values, converted to redshift space
and divided by the corresponding linear bias factor.
}
         \label{fig:s4}
   \end{figure*}

In Figs. \ref{fig:s3} and \ref{fig:s4} we show $S_3$ and 
$S_4$ as a function of luminosity and stellar mass in the three redshift bins.
We also show the predictions of second--order perturbation theory in real space
for the matter distribution and the corresponding predictions for galaxies, 
derived from the matter value assuming the linear bias estimated from 
$\overline{\xi_2}$, and corrected for redshift space distortion using
the factors obtained from mocks. This derivation is described in the next 
subsection.
The theoretical curves for $S_3$ and $S_4$ are shown for radii larger than 
$\sim 6 h^{-1} $ Mpc, as they are calculated in the quasi--linear regime.


In the first redshift bin, both for luminosity and stellar mass limited 
samples, the value of $S_3$ is constant and around $2$ at small and 
intermediate scales, but it starts decreasing beyond $R \sim 8 h^{-1}$ Mpc.
In principle, variations of $S_3$ with scale can be due to changes in the 
slope of the power spectrum or to a scale--dependent bias. However, such a 
systematic effect can be ascribed to the small number of independent cells at 
large scales, as shown by mocks and reflected in the large error bars.
In the same redshift bin, $S_4$ shows a small decrease at large scales and
is consistent with a constant value of $\sim 7.3$ between 4 and 10 $h^{-1}$ Mpc.
In the range $6-8 h^{-1}$ Mpc, the best scales to compare with
perturbation theory (on larger scales the errors increase significantly), 
the theoretical predictions for $S_3$ and $S_4$ 
are slightly higher than the observed values
corresponding to the $M_B(z=1.1)-5\log(h) \le -20.5$ subsample, but
 only at $1 \sigma$ level.

In the second redshift bin the value of $S_3$ for luminosity--limited 
subsamples is around $1.8$, sligthly lower than in the first bin, 
but still consistent within the errors; moreover, it is 
consistent with a constant value in the whole range of scales. 
The value of $S_3$ for mass--limited 
subsamples is also constant in the whole range of scales and is consistent with the value in the first redshift bin.
$S_4$ has an analogous behaviour: while showing a systematic decrease, 
particularly in luminosity--limited subsamples, it is still consistent with
a constant value in the range $4-16 h^{-1}$ Mpc.
As in the case of the first redshift bin, in the range $6-8 h^{-1}$ Mpc
the theoretical predictions for $S_3$ and $S_4$ 
are slightly higher than the corresponding observed values.

In the third redshift bin the values of $S_3$ and $S_4$ for luminosity- and
 stellar-mass-limited subsamples increase systematically with scale.
Moreover, in contrast with the two previous redshift bins, 
 in the range $6-8 h^{-1}$ Mpc, 
the theoretical predictions for $S_3$ and $S_4$ 
are {\em \textup{lower}} than the observed values.

To better appreciate the significance of these deviations, 
we note that of 26 mocks, 3 show an increase of the values of 
$S_3$ and $S_4$ similar to what we find in the last redshift bin.

In fact, 
higher--order statistics are very sensitive to large--scale structure, and 
the correlated variations in the measured values of $S_3$ and $S_4$ 
probably indicate genuine fluctuations in the galaxy distribution
(see e.g. the discussion in \citealp{2004MNRAS.352.1232C}). 

In our case, this interpretation is suggested by checking the 
$W1$ and $W4$ fields separately: we find that in the outermost redshift shell, 
both $S_3$ and $S_4$ are larger in $W1$  than in $W4$. 
For example, for the $M_B(z=1.1)-5\log(h) \le -21.0$ subsample,
at R=8 $h^{-1}$ Mpc, we find $S_3 = 2.7 \pm 0.5$ in $W1$ and 
$S_3=1.6 \pm 0.3 $ in $W4$.
Analogously, for the $\log(M/M_\odot h^{-2}) \ge 10.5$ subsample
at R=8 $h^{-1}$ Mpc, we find 
$S_3 = 3.4 \pm 0.5$ in the $W1$ field and $S_3=2.0 \pm 0.3$ in the $W4$ field. 
This difference might be regarded as the imprint of spatially coherent 
structures more prominent in $W1$.

In conclusion, 
the values of $S_3$ and $S_4$ do not show any significant dependence 
on luminosity or on stellar mass:
the points corresponding to different subsamples are consistent within the 
error bars (we discuss a possible weak 
dependence on luminosity in the next subsection).
There is no evidence of evolution in redshift either,
apart from the systematic increase of $S_3$ and $S_4$ with scale
in the last redshift bin.

Taking into account the behaviour of mocks,
the observed systematic variations in the values of high--order 
moments are consistent with the fluctuations expected for comparable
volumes randomly extracted from a $\Lambda$CDM universe.

It is possible to compare our results on $S_3$ and $S_4$ with those obtained
by \cite{2013MNRAS.tmp.2056W} 
for the four CFHTLS-Wide 
fields. They have  divided the galaxies in the photometric catalogue into four 
redshift bins through the estimated photometric redshifts; 
for galaxies with $M_g < -20.7$, they have estimated $S_J$ as a function of 
angular scale and the 
corresponding $3D$ values through deprojection, which, as they discussed, rely on 
some approximations. 
Their work is therefore complementary to ours: they have a larger
area and number of objects, but we can directly estimate the 3D (redshift 
space) $S_J$; they can sample smaller, highly non--linear scales where we
do not have enough statistics, but we can better sample the quasi--linear 
scales; finally, we can also test the dependence of 
$S_3$ on luminosity and stellar mass.

A comparison with their Fig. 12 shows that, as expected 
(see our Fig.\ref{fig:realspace}), their deprojected 
values for $S_3$ and $S_4$ on small scales ($R  < 5$ $h^{-1}$ Mpc) are higher 
than our redshift space values. On larger scales, the redshift space effect on 
$S_3$ and $S_4$ becomes negligible, and their estimate is consistent
with ours.

We note that \cite{2013MNRAS.tmp.2056W} 
found significant deviations in the results for the $W3$ field,
while we have found differences between $W1$ and $W4$ in our last redshift bin:
this shows that sample variance is
still significant for high--order statistics on the scale of CFHTLS Wide Fields.

%
%
\subsection{Implications for biasing}

We now discuss the implications of our analysis for biasing. 
We concentrate on the reference scale $R=8 h^{-1}$ Mpc, 
where second--order perturbation theory predictions can be applied and 
results are still reliable (errors and systematic 
deviations increase on larger scales).
Because we aim to compare our results with the matter density field, 
statistical quantities referring to galaxies are indicated 
with a subscript $g$ and those relative to matter with a subscript
$m$.  

\begin{figure}
   \centering
       \resizebox{\hsize}{!}{\includegraphics[width=17cm]{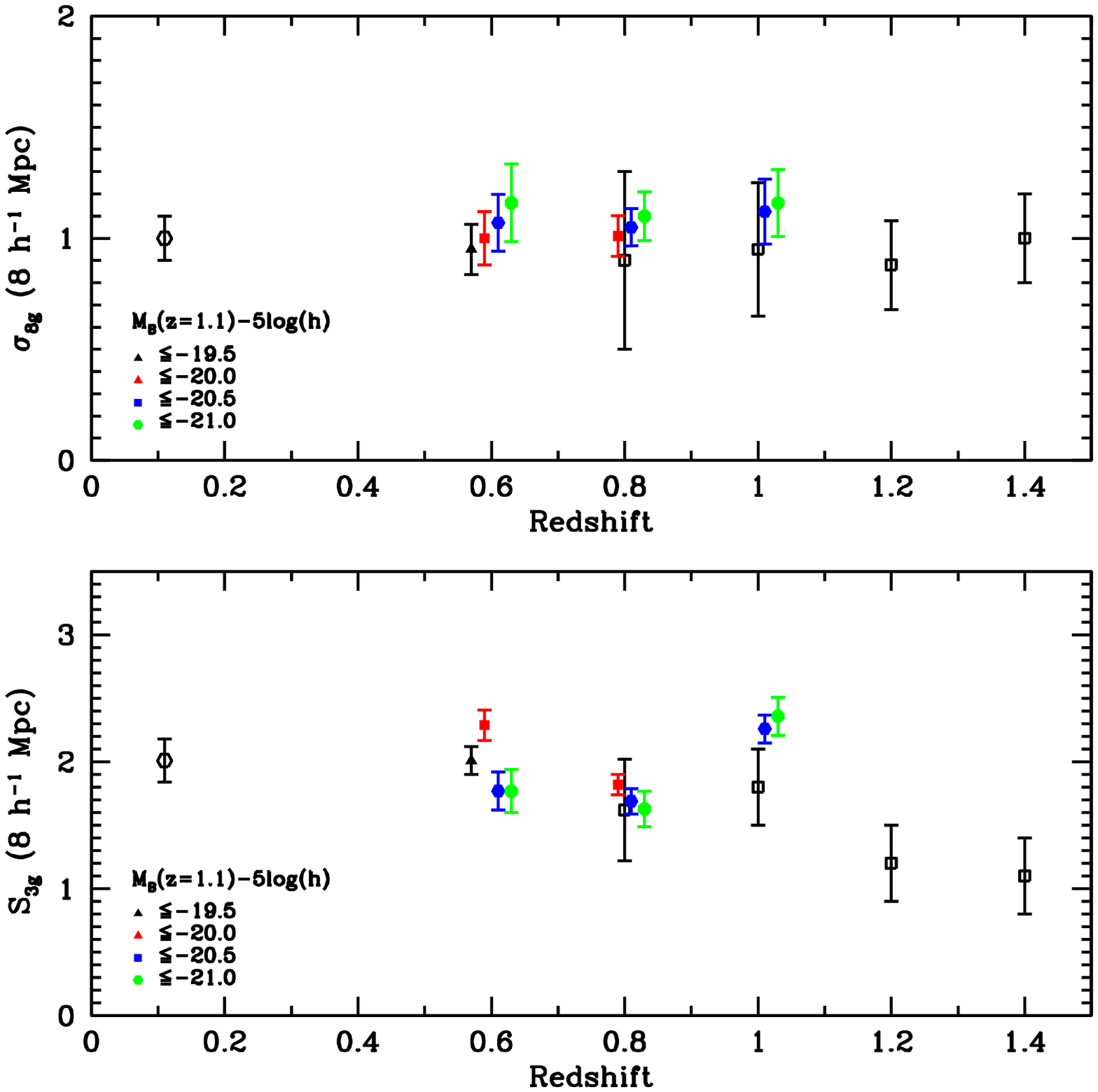}}
      \caption{$\sigma_{8g}$ (upper panel) and $S_{3g}$ (lower panel) as a 
function of redshift. 
Colours represent the different absolute magnitude limits as in previous figures.
The open hexagon represents the corresponding value for 2dFGRS galaxies in the
absolute magnitude $M_{b_J}$ range [-21,-20] 
(\citealp{2004MNRAS.352.1232C}). 
Black open squares are the values for VVDS galaxies
brighter than $M_B =-21$ 
(\citealp{2005A&A...442..801M}). 
For better visibility, points corresponding to different
redshifts are slightly shifted in magnitude.}
         \label{fig:s83z}
   \end{figure}

Figure \ref{fig:s83z} shows the values of $\sigma_{8g}$ (top panel) and $S_{3g}$ 
(bottom panel) at $R=8 h^{-1}$ Mpc for the VIPERS volume--limited 
subsamples with different limiting absolute magnitudes and in the different 
redshift bins.
In the same figure we also show the corresponding VVDS estimates 
(\citealp{2005A&A...442..801M}) 
and the 2dFGRS estimates for
the local Universe (\citealp{2004MNRAS.352.1232C})  
for galaxies with a similar luminosity as ours.

At a given redshift, VIPERS subsamples with a brighter absolute magnitude limit 
have higher values of $\sigma_{8g}$,
but there is no significant evolution of $\sigma_{8g}$ with redshift.
The same holds when combining our results with those of the 2dFGRS
in the local Universe and those of the VVDS at higher redshift: 
$\sigma_{8g}$ shows no significant evolution from $z=0$ to $z=1.4$ 
(VVDS points are systematically lower but at the
$1\sigma$ level).
This implies (see e.g. the discussion in 
\citealp{2005A&A...442..801M}) 
a strong evolution of the
linear bias $b$ with redshift 
because $\sigma_{8m}$ increases with time (see Eq. 
\ref{eq:bias}).
There are various models that describe the evolution of $b(z)$ and
explain its decrease with time (see e.g.\citealp{2000ApJ...531....1B});
from an empirical point of view, we note that the available data can be fitted 
by the simple relation $b(z) \propto 1 / \sigma_{8m}$. 

The skewness $S_{3g}$ of the VIPERS subsamples measured at 8 $h^{-1}$ Mpc 
and plotted as a function of redshift has more fluctuations than
$\sigma_{8g}$, with a minimum value in the redshift bin [0.7,0.9], but it 
does not show a significant dependence on luminosity and is still consistent
with a constant value independent of redshift. 
The values of $S_{3g}$ in the VVDS below 
$z=1.2$ are lower than VIPERS values, but are consistent within the errors,
while they start to decrease beyond $z \sim 1.1$.

The absence of a significant evolution of $S_{3g}$ with redshift is not limited
to our redshift range: the values of $S_{3g}$ measured in VIPERS
are similar to those measured in the 2dFGRS, that is,
 $S_3 \sim 2.0 \pm 0.2$, where depending on the subsample $S_{3g}$ varies 
from 1.95 to 2.58 (while not shown in the figure, the values of $S_4$ are 
also consistent with the 2dFGRS ones).
Therefore, taking into account all data points, starting from the local value
for the 2dFGRS up to $z=1.1$ (VIPERS and VVDS data), $S_{3g}$ is consistent
with a constant value $\sim 2$: in VIPERS the strongest but marginal
deviations of the $S_{3g}$ value are for $M_B(z=1.1)-5\log(h) \le -20.0$ 
galaxies in the nearest redshift range [0.5,0.7] and for
 $M_B(z=1.1)-5\log(h) \le -21.0$ galaxies in the most distant redshift interval
 [0.9,1.1], both giving a value of $S_{3g}$ that is 15\% higher.

%
%

\begin{figure}
   \centering
       \resizebox{\hsize}{!}{\includegraphics{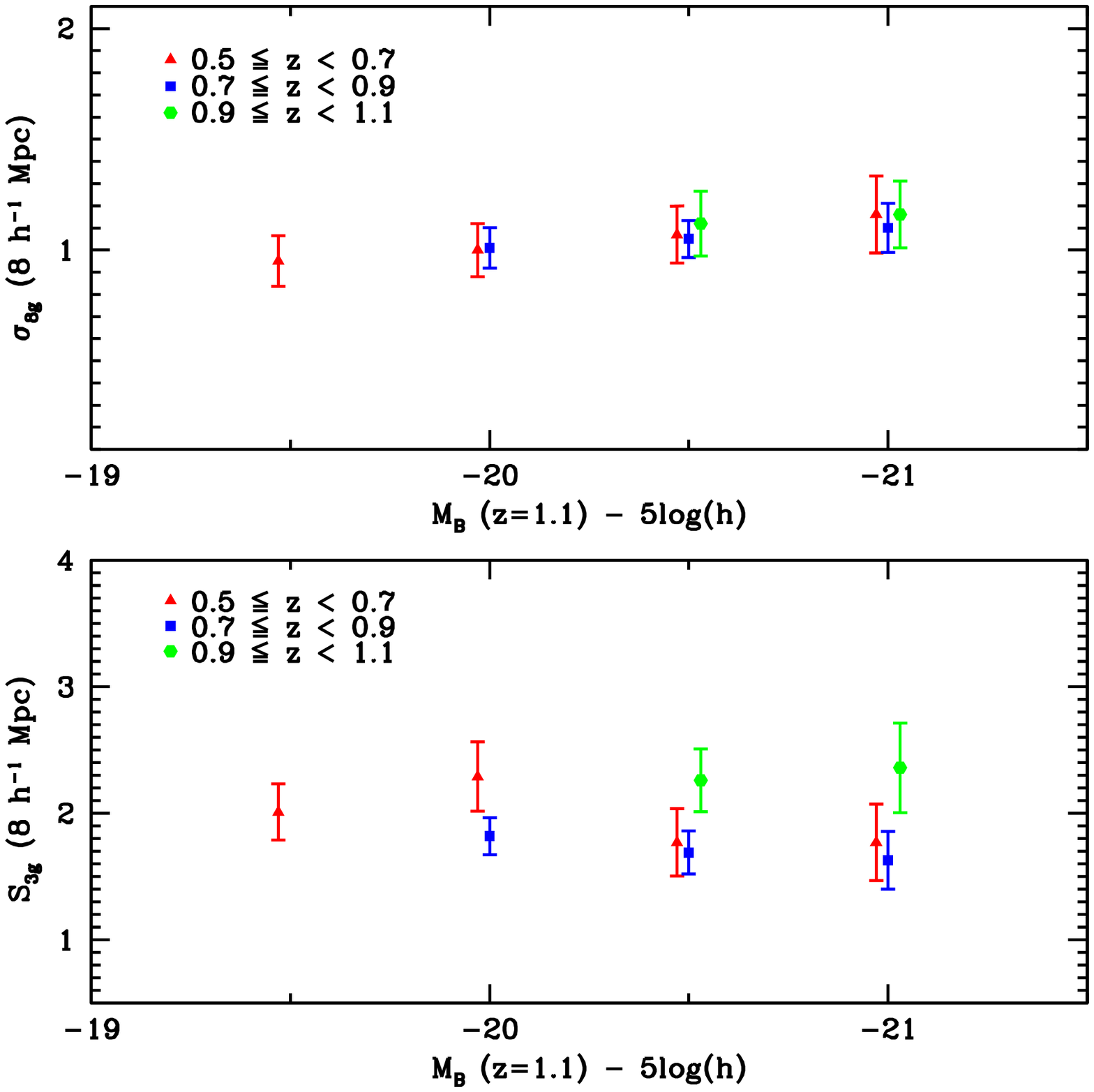}}
      \caption{$\sigma_{8g}$ (upper panel) and $S_{3g}$ at R=8 $h^{-1}$ Mpc 
(lower panel) 
as a function of galaxy luminosity for the three redshift bins. Red triangles: 
$0.5 \le z < 0.7$; blue squares: $0.7 \le z < 0.9$; green hexagons: 
$0.9 \le z < 1.1$. For better visibility, points corresponding to different
redshifts are slightly shifted in magnitude.}
         \label{fig:sigma8}
   \end{figure}

Figure \ref{fig:sigma8} shows $\sigma_{8g}$ (top panel) and $S_{3g}$ 
at $8 h^{-1}$Mpc 
(bottom panel) as a function of absolute magnitude for the three 
redshift bins. 
$\sigma_{8g}$ shows a systematic increase with luminosity
(reflecting the dependence of the correlation amplitude on
luminosity), but at a given absolute 
luminosity its value is similar in the three redshift bins. 

$S_{3g}$ appears to be independent of absolute magnitude, 
with fluctuations from sample to sample. However, 
if we exclude the points relative to the last redshift bin, 
where  $S_{3g}$ has a higher value, the
data might suggest a small decrease of $S_{3g}$ with increasing luminosity, 
reminiscent of the results of \cite{2004MNRAS.352.1232C} 
for the 2dFGRS.

A trend of $S_{3g}$ with luminosity is
interesting because in the hypothesis of linear biasing, 
$S_{3g}$ is inversely proportional to the bias factor $b$: 
knowing from the two--point correlation function of our samples 
that $b$ increases with luminosity, we expect a corresponding decrease of $S_{3g}$.

%
%

To test whether our results are consistent with the linearity 
of bias, we therefore estimated the bias of galaxies with respect to the 
underlying 
matter density field at $R=8 h^{-1}$ Mpc, using the observed
$\sigma_{8g}$ and $S_{3g}$ of the galaxy distribution and estimating $\sigma _m$
and $S_{3m}$ of the matter distribution through perturbation theory. 

\cite{1993ApJ...412L...9J} 
and \cite{1994ApJ...433....1B,1994A&A...291..697B}
(see also \citealp{2002PhR...367....1B} 
and references therein) have shown that for a smoothed density field 
with primordial Gaussian fluctuations, Peebles' unsmoothed value of 
$S_{3m} = 34/7$ (\citealp{1980lssu.book.....P}) 
has to be corrected according to the expression
\begin{equation}
 S_{3m} = 34/7 + d\ln \sigma_m ^2 / d \ln R ,
\label{eq:s3ps}
\end{equation}

where $d\ln \sigma_m ^2 / d \ln R$ is the
logarithmic slope of the linear variance of the matter density field smoothed 
with a spherical top--hat function of radius $R$,

\begin{equation}
 \sigma_m ^2(R) = \frac {1}{2 \pi^2} \int _0 ^{\infty} dk k^2 P(k) W^2(kR) .
\label{eq:sigma2}
\end{equation}

For a power--law spectrum $P(k) \propto k^n$, Eq. \ref{eq:s3ps} becomes
$S_{3m} = 34/7 - (n+3)$.

Similar relations hold for higher orders, involving higher--order 
derivatives. 

%
The values obtained from perturbation theory have been tested with numerical 
simulations, and it has been shown that in the
range we are studying, that is, at $R=8 h^{-1}$ Mpc and for $\sigma_{8m} \sim 1$,
they are very accurate: for example, the difference in the $S_3$ values 
is smaller than a few percent 
(\citealp{1995MNRAS.274.1049B}, 
\citealp{1998MNRAS.301..503F},  
\citealp{2002PhR...367....1B}). 

Applying Eqs. (\ref{eq:s3ps}) and (\ref{eq:sigma2}) and using the 
software {\small CAMB} (\citealp{2002PhRvD..66j3511L}), 
we have computed the values of $\sigma_{8m}$ and $S_{3m}$ for a power spectrum 
with the new cosmological parameters derived from the 
Planck mission (\citealp{2013arXiv1303.5076P}) 
and with the old Millennium parameters (first year WMAP data and 2dFGRS,
with $\Omega_M=0,25$, $\Omega_\Lambda=075$, $n=1$ and $\sigma_{8m}=0.9$).

%
We here assumed that the standard $\Lambda CDM$ model is correct.
With other assumptions, such as a dark energy component with an evolving 
equation of state or modified gravity, the clustering and bias evolution
would be affected (see e.g. \citealp{2004MNRAS.349..281M}), 
as would the redshift distortions (\citealp{2013MNRAS.435.2806H}). 
This dependence on cosmology will be studied in a future work.

We also converted the observed $\sigma_{8g}$ and $S_{3g}$ to real space 
values 
by applying correction factors directly derived from the mocks.

For the subsample limited at $M_B (z=1.1) - 5 \log(h) \le -20.50$, 
 we give in Table 2 the redshift range (Col. 1),
the values of $\sigma_{8g}$ (Col. 2), $\sigma_{8m}$ (Col. 3), 
$b=\sigma_{8g}/\sigma_{8M}$ (Col. 4),
$S_{3g}$ (Col. 5), $S_{3m}$ (Col. 6), 
all measured at a scale of $R=8 h^{-1}$ Mpc.

%
%

\begin{table*}
\caption{Real space values of $\sigma_8$ and $S_3$ of galaxies with 
$M_B (z=1.1) - 5 \log(h) \le -20.50$ 
vs. those expected from second--order perturbation theory 
(with Millennium and Planck cosmological parameters).}     
\label{table:2}     
\centering                                     
\begin{tabular}{c c c c c c}          
\hline\hline                        
Redshift range     & $\sigma_{8g}$   & $\sigma_{8m}$        & $b$            & $S_{3g}$ ($R = 8 h^{-1}$ Mpc) & $S_{3m}$ ($R = 8 h^{-1}$ Mpc) \\
                   & (real space)    & (WMAP/Planck) & (linear bias)   & (real space)          & (WMAP/Planck) \\
\hline 
$0.5 \le z < 0.7$  & $1.00 \pm 0.12$ & 0.61 / 0.63         & $1.65 \pm 0.20$ & $1.91 \pm 0.29$       &  3.52 / 3.48        \\
$0.7 \le z < 0.9$  & $0.98 \pm 0.08$ & 0.55 / 0.57         & $1.77 \pm 0.14$ & $1.76 \pm 0.18$       &  3.52 / 3.48        \\
$0.9 \le z < 1.1$  & $1.04 \pm 0.13$ & 0.51 / 0.52         & $2.04 \pm 0.27$ & $2.28 \pm 0.25$       &  3.52 / 3.48        \\
\hline                                   
\end{tabular}
\end{table*}

\begin{figure}
   \centering
       \resizebox{\hsize}{!}{\includegraphics{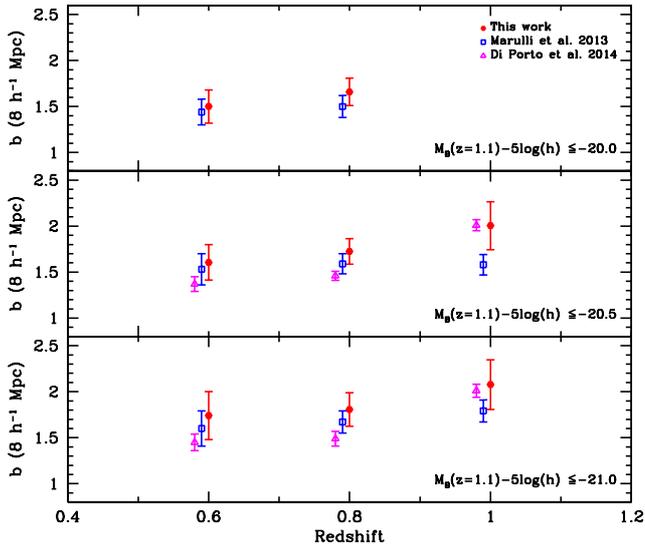}}
      \caption{Linear bias $b$ as a function of redshift.
Top panel: $M_B \le -20.0(z=1.1) +5 \log(h)$; middle panel: 
$M_B \le -20.5(z=1.1) +5 \log(h)$; 
bottom panel: $M_B \le -21.0(z=1.1) +5 \log(h)$. 
Red hexagons: our estimates of $b = \sigma_{8g}/\sigma_{8m}$.
Blue squares: estimates of 
\cite{2013A&A...557A..17M}. 
Magenta triangles: 
 \cite{2014arXiv1406.6692D}. 
}
         \label{fig:bz}
   \end{figure}

In Fig. \ref{fig:bz} we plot our estimates for the linear bias term $b$
as a function of redshift, with the correponding estimates for VIPERS of 
\cite{2013A&A...557A..17M} 
and \cite{2014arXiv1406.6692D}. 
As expected, the estimates are fully consistent, with
$b$ increasing with luminosity and redshift.
 As discussed by 
 \cite{2014arXiv1406.6692D}, 
there is only a difference in the last redshift bin
where the estimate of Marulli et al. is lower than that of 
 \cite{2014arXiv1406.6692D}. 
The difference is probably due to the way $b$ is
estimated (counts in cells in our case and in
\citealp{2014arXiv1406.6692D}, 
pair counts in \citealp{2013A&A...557A..17M}). 
Our estimate is consistent with both the other two estimates
at the $1 \sigma$ level, however.

%
%

\begin{figure}
   \centering
       \resizebox{\hsize}{!}{\includegraphics{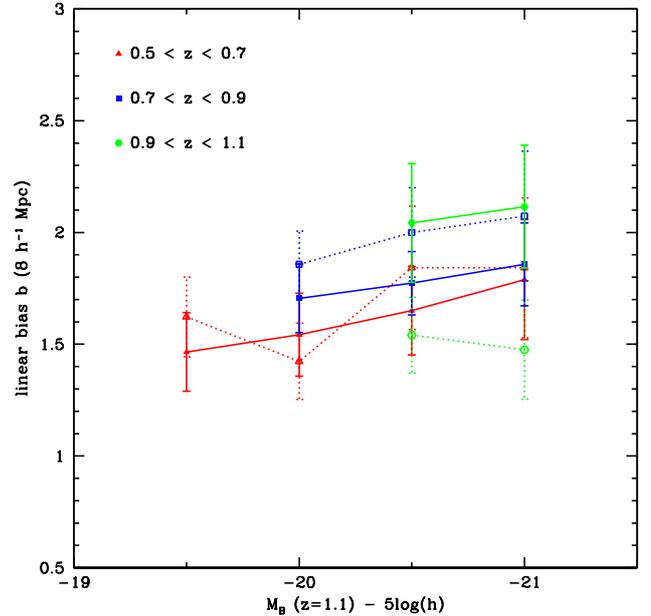}}
      \caption{Estimates of $b = \sigma_{8g}/\sigma_{8m}$ (solid lines, 
filled symbols) and $b^\prime \equiv S_{3m}/S_{3g}$ (dotted lines, open symbols); 
$b^\prime = b$ under the assumption of linear bias. 
Red lines with triangles: $0.5 \le z < 0.7$; 
blue lines with squares: $0.7 \le z < 0.9$; 
green lines with hexagons: $0.9 \le z < 1.1$.}
         \label{fig:linearbias}
   \end{figure}

In Fig. \ref{fig:linearbias} we compare the linear bias directly measured from
the ratio of the galaxy and matter rms, $b = \sigma_{8g} / \sigma_{8m}$, 
with the ratio of the galaxy and matter skewness, $S_{3m} /  S_{3g}$. 
Under the hypothesis of linear biasing, the two ratios should have the same
value.
For the first two redshift bins we find slightly different values:
the skewness ratio is systematically
higher than the bias directly computed from the variance. 
The third redshift bin shows the largest discrepancy, but with the opposite 
behaviour, that is, the skewness ratio is lower 
than the bias directly computed from the variance.  
This different behaviour is a consequence of the fact that the value of 
$S_{3g}$ in the last redshift bin increases with scale and becomes higher than 
at lower redshifts.

We can quantify the degree of non--linearity by directly estimating 
the second--order term $b_2$ from Eq. \ref{eq:secondorderbias}: 

\begin{equation}
 b_2 = b (b S_{3g} - S_{3m})/3,\end{equation}


where we used the real space values $S_{3g}$ and $b$ obtained from the
redshift space values by using the conversion factor calculated from the mocks.
We note that this correction is small (a few percent) at our scale of 
$R=8 h^{-1}$ Mpc, because this scale is at the 
transition from the regime of small-scale velocity dispersion (where 
redshift space correlation functions are lower than real space ones) to the 
regime of infall where redshift space correlations are higher than real space 
ones (see Fig. \ref{fig:realspace}).

In this formalism, if $b > 0$, $b_2$ is negative when 
$\sigma_{8g}/\sigma_{8m} < S_{3m}/S_{3g}$. This is what happens in the first
two redshift bins, where at nearly all magnitudes $b_2$ is negative:
for example, for the subsample limited at $M \le -20.5$(z=1.1) - 5 $\log$(h), 
we find $b_2 = -0.20 \pm 0.49$ in the first redshift
bin and  $b_2 = -0.24 \pm 0.35$, in the second redshift bin. 
In contrast, we find a positive $b_2$ in the
third bin,
with $b_2 = +0.78 \pm 0.82$. 


As we have noted above when discussing the results of our tests on mocks,
the assumption that masked regions and inhomogeneities
can be described as a Poissonian random sampling gives a small bias
with an overestimate of $b$ of a few percent and an underestimate of $S_3$ 
around 10-15\%.
Using the correction factors derived from the average of the 
mocks, we find for the subsample limited at $M \le -20.5$(z=1.1) - 5 $\log$(h) 
$b_2 = -0.03 \pm 0.49$ in the first redshift, 
$b_2 = -0.25 \pm 0.35$ in the second redshift bin, and
$b_2 = +0.72 \pm 0.82$ in the third bin.
The differences are well within $1 \sigma$ error.

It would be tempting to interpret these results as suggesting a possible 
evolution of the non--linear bias $b_2$ with redshift, with a similar trend, for example, as for the model of
\cite{2007PhRvD..76h3004S}. 
Unfortunately, the problem is the extreme sensitivity of $b_2$ to the
errors on $b$ and $S_{3g}$, amplified by a factor $b^2$, and we have seen that
subsamples in the last redshift bin are affected by larger errors and
systematic trends.

With these caveats, we can check the consistency of our results with 
other works in the same redshift and luminosity ranges. 
In this comparison,
one has to take into account the sensitivity of $b_2$ to the 
different methods and, as pointed out by \cite{2011ApJ...731..102K}, 
to sample variance. In fact, even local measurements of the non--linear term 
have 
given different values 
(see e.g. \citealp{2002MNRAS.335..432V} 
and \citealp{2005MNRAS.362.1363P}, and the discussion in 
\citealp{2005MNRAS.364..620G} and \citealp{2008A&A...487....7M}).

First of all, \cite{2014arXiv1406.6692D} 
have analysed the VIPERS data  
reconstructing the bias relation from the estimate of the 
probability distribution function: they found 
a small (< 3\%) but significant deviation from linear bias.

In their analysis of the four CFHTLS Wide fields, 
\cite{2013MNRAS.tmp.2056W} 
have found that perturbation theory predictions agree well with
their measurements when taking into account the linear bias, but note that there is
still a small discrepancy that can be explained by the presence of
a non--linear bias term. This is also consistent with what we found.

\cite{2005A&A...442..801M} 
have analysed VVDS volume--limited samples limited at $M_B < -20 + 5 \log(h)$
(this limit was fixed and did not take into account luminosity
evolution) in the redshift bins $0.7 < z < 0.9$ and $0.9 < z < 1.1$,
finding $b_2 = -0.20 \pm 0.08$ and 
$b_2 = -0.12 \pm 0.08$ (here the errors do not include sample variance): 
these values are consistent with ours below $z=1$.

\cite{2011ApJ...731..102K} 
analysed the zCOSMOS galaxy overdensity field and estimated the mean 
biasing function between the galaxy and matter density fields and its second 
moment, finding a small non--linearity, with the nonlinearity parameter
$\tilde{b}/\hat{b}$ (defined in the formalism of 
\citealp{1999ApJ...520...24D}) 
at most 2\% with an uncertainty of the same order.

\cite{2005MNRAS.364..620G} have found 
$b= 0.93 ^{+0.10} _{-0.08}$ and $c_2=b_2/b=-0.34 ^{+0.11} _{-0.08}$
from the measurement of the $Q_3$ parameter in the three--point correlation 
function of the 2dFGRS for the local Universe.

The non--linear term we have measured in the redshift interval 
between $z=0.5$ and $z=0.9$ is therefore similar to what has been 
measured in the above surveys.
We conclude that there is general evidence for a small but non--zero 
non--linear $b_2$ term.
It is also clear that no evolution of $b_2$ with redshift can be detected 
in the available data, in contrast to the linear bias term.

%
%
\section{Conclusions}

We have analysed the high--order clustering of galaxies 
in the first release of VIPERS, using counts in cells to derive the 
volume--averaged correlation functions and normalized skewness $S_{3g}$ and 
kurtosis $S_{4g}$. We have analysed volume--limited subsamples with 
different cuts in absolute magnitude and stellar mass in three redshift bins;
these subsamples are the same as in
\cite{2013A&A...557A..17M}. 

Errors were estimated through a set of mock catalogues, derived from dark matter
halo catalogues repopulated with the method of 
\cite{2013MNRAS.435..743D}. The mocks were built to
reproduce the properties of VIPERS, including masks and selection effects.
Our analysis has shown that the high--order statistical properties of these
mocks are consistent with observations.

We also studied the dependence of the second-- and third--order 
statistics of
galaxy counts on the bias, deriving the linear bias term $b$ and 
the first non--linear term $b_2$, and comparing our results with predictions from 
perturbation theory and with other works in the literature. 

Here are our main conclusions. 

\begin{itemize}

\item We showed that the hierarchical scaling relations 
$\overline{\xi}_3 \propto \overline{\xi}_2 ^2$ and 
$\overline{\xi}_4 \propto \overline{\xi}_2 ^3$ 
hold in the range of scales and redshifts we could sample, that is, 
$3 \le R \le 10 h^{-1}$ Mpc and $0.5 \le z < 1.1$. 
These relations are consistent with
predictions from gravitational clustering and with the scaling observed in local
surveys.

\item $S_{3g}$ and $S_{4g}$ appear to be independent of luminosity; 
however, if we do not not take the last redshift bin into account,
there is a slight decrease of $S_{3g}$ with increasing luminosity, 
an effect previously detected locally in the 2dFGRS by 
\cite{2004MNRAS.352.1232C}. 

\item The values of $S_{3g}$ and $S_{4g}$ are scale--independent within
 the errors and do not evolve significantly at least up to $z=0.9$.
We detected a systematic increase with scale in the last redshift 
bin  (beyond  $\sim 10 h^{-1}$ Mpc), mainly due to one of the two CFHTLS 
fields ($W1$); this  deviation is consistent with 
what can be expected from the sample variance shown by mock catalogues.

\item The observed values of $S_{3g} \sim 2 \pm 0.2$ and 
$S_{4g} \sim 8 \pm 0.4$ are
similar to those measured in local surveys for galaxies in the same
luminosity range. This confirms the substantial absence of evolution of
$S_{3g}$ in the redshift range $0 < z < 1$ at the level of $\sim 10$\%.
This result is expected for $S_{3m}$ , but is not trivial
for $S_{3g}$, given the evolution of bias.

\item At second order, galaxies with higher luminosity or stellar mass 
have a larger amplitude (greater linear bias parameter) of the volume--averaged 
two--point correlation function, consistently with the direct analysis of the 
two--point correlation function by 
\cite{2013A&A...557A..17M}. 
We  showed that our estimate of the linear bias parameter 
$b=\sigma_{8g}/\sigma_{8m}$ is consistent 
within $1 \sigma$ with those of \cite{2013A&A...557A..17M} 
and \cite{2014arXiv1406.6692D}. 
The linear bias increases both with luminosity and with redshift:
in our redshift range, we measured the lowest bias $b=1.47 \pm 0.18$ for galaxies
 with  $M_B(z=1.1)-5\log(h) \le -19.5$  in the redshift bin 
$0.5 \le z < 0.7$ and the 
largest bias  $b=2.12 \pm 0.28$ for galaxies with 
$M_B(z=1.1)-5\log(h) \le -21.0$ in the redshift bin $0.9 \le z < 1.1$.

\item For a given luminosity class, $\sigma_{8g}$ does not evolve with 
redshift. For example, comparing our values for $M_B(z=1.1)-5\log(h) \le -20.5$
to the corresponding value measured in the 2dFGRS, we found that $\sigma_{8g}$
is consistent with a constant value ~1.0 (our $1 \sigma$ error is 10\%), 
from $z=0$ to $z \sim 1$. Given that $\sigma_{8m}$ increases with time,
we have the empirical relation $b(z) \propto 1/ \sigma_{8m}(z)$.

\item The value of the non--linear bias parameter $b_2$ measured below 
$z \sim 1$ at the scale $R = 8 h^{-1}$ Mpc, that is, in the quasi--linear regime, 
is negative but not statistically
different from zero when taking into account the error; however, taking into 
account
the ensemble of results coming from this and other surveys in the redshift 
range $0.5 \le z < 1$
(\citealp{2005A&A...442..801M}, 
 \citealp{2011ApJ...731..102K}, 
 \citealp{2013MNRAS.tmp.2056W}, 
 \citealp{2014arXiv1406.6692D}), 
there is evidence for a small but non--zero non--linear term.
Including the results from local surveys as well, no evolution of 
$b_2$ with redshift can be detected in the available data.

\item The comparison with the properties of mocks and with the predictions of 
perturbation theory shows that our results are consistent with the general 
scenario of biased galaxy formation and gravitational clustering evolution in 
a standard $\Lambda CDM$ cosmology.

\end{itemize}

In conclusion, we have provided an independent check 
on the second--order statistical studies of the galaxy distribution
through our analysis; we 
explored the galaxy bias with an independent technique; 
finally, we determined the higher--order statistical properties of the 
galaxy distribution in the redshift range between 0.5 and 1.1, 
thanks to the combination of volume and density of galaxies in the 
VIPERS survey.
When VIPERS is complete, it will be possible to perform a more 
general analysis, which will allow us not only to decrease
error bars, but also to include the dependence of high--order statistics on 
galaxy colour, to apply other high--order statistical tools such as the void 
probability function, and to give better constraints on the non--linear bias.

%
%

\begin{acknowledgements}

This work is based on observations collected at the European
Southern Observatory, Cerro Paranal, Chile, using
the Very Large Telescope under programs 182.A-0886
and partly 070.A-9007. Also based on observations obtained
with MegaPrime/MegaCam, a joint project of CFHT
and CEA/DAPNIA, at the Canada-France-Hawaii Telescope
(CFHT), which is operated by the National Research Council
(NRC) of Canada, the Institut National des Sciences de l’Univers
of the Centre National de la Recherche Scientifique (CNRS) of
France, and the University of Hawaii. This work is based in
part on data products produced at TERAPIX and the Canadian
Astronomy Data Centre as part of the Canada-France-Hawaii
Telescope Legacy Survey, a collaborative project of NRC and
CNRS. The VIPERS web site is http://www.vipers.inaf.it/. We
acknowledge the crucial contribution of the ESO staff for the
management of service observations. In particular, we are deeply
grateful to M. Hilker for his constant help and support of this
program. Italian participation to VIPERS has been funded by
INAF through PRIN 2008 and 2010 programs. DM gratefully
acknowledges financial support of INAF-OABrera. LG, AJH,
and BRG acknowledge support of the European Research Council
through the Darklight ERC Advanced Research Grant (\#291521). 
AP, KM, and JK have been supported by the National
Science Centre (grants UMO-2012/07/B/ST9/04425 and
UMO-2013/09/D/ST9/04030), the Polish-Swiss Astro Project
(co-financed by a grant from Switzerland, through the Swiss
Contribution to the enlarged European Union), and the European
Associated Laboratory Astrophysics Poland-France HECOLS.
KM was supported by the Strategic Young Researcher Overseas
Visits Program for Accelerating Brain Circulation (\#R2405).
OLF acknowledges support of the European Research Council
through the EARLY ERC Advanced Research Grant (\#268107). 
GDL acknowledges financial support from the European 
Research Council under the European Community’s
Seventh Framework Programme (FP7/2007-2013)/ERC grant
agreement \# 202781. WJP and RT acknowledge financial support
from the European Research Council under the European
Community’s Seventh Framework Programme (FP7/2007-
2013)/ERC grant agreement \#202686. WJP is also grateful for
support from the UK Science and Technology Facilities Council
through the grant ST/I001204/1. EB, FM and LM acknowledge
the support from grants ASI-INAF I/023/12/0 and PRIN MIUR
2010-2011. LM also acknowledges financial support from PRIN
INAF 2012. YM acknowledges support from CNRS/INSU (Institut
National des Sciences de l’Univers) and the Programme
National Galaxies et Cosmologie (PNCG). CM is grateful for
support from specific project funding of the Institut Universitaire
de France and the LABEX OCEVU. SdlT acknowledges the support of the 
OCEVU Labex (ANR-11-LABX-0060) and the A*MIDEX project (ANR-11-IDEX-0001-02) 
funded by the "Investissements d'Avenir" French government program managed 
by the ANR. 
\end{acknowledgements}

%
%

\bibliographystyle{aa}
\bibliography{References}

\end{document}